\documentclass[11pt,a4paper]{article}
\usepackage{geometry}  
\usepackage[british]{babel} % [british]
\usepackage[T1]{fontenc}
\usepackage[utf8]{inputenc}
\usepackage[pdftex]{graphicx} % or [dvips]
\usepackage{url}
\usepackage{color}
\usepackage{fourier,charter}
\usepackage{varioref}\labelformat{equation}{(#1)}
\usepackage[round]{natbib}
\usepackage{enumerate,mathrsfs}
\title{Some computational aspects of maximum likelihood estimation 
       of the skew-$t$ distribution}
       
\author{{\Large Adelchi Azzalini} \\  \large 
  Dipartimento di Scienze Statistiche \\
  Università di Padova, Italia
  \and
  {\Large Mahdi Salehi}\\
  Department of Mathematics and Statistics\\
   University of Neyshabur, Iran
 }
\date{16th June 2019}
\newcommand{\E}[2][]{
   \ensuremath{\mathbb{E}_{#1}\!\left\{\displaystyle{#2}\right\}}}

\newcommand{\Real}{\mathbb{R}}
\newcommand{\half}{\mbox{$\textstyle \frac{1}{2}$}}  
\newcommand{\inv}{^{-1}}
\newcommand{\T}{^{\top}}
\newcommand{\ST}{\mathrm{ST}}
\newcommand{\negindent}{\par\vskip0.5ex\noindent\hangindent=2em\hangafter=1}
\newcommand{\subZ}{_{\tinyZ}} % assume math-mode
\newcommand{\tinyZ}{\mbox{\tiny\textit{Z}}} % assume math-mode
\newcommand{\R}{\texttt{R}}
\usepackage{placeins}  %for FloatBarrier

\begin{document}
\maketitle

%------------------------------------------------------------------------------

\abstract{Since its introduction, the skew-$t$ distribution has received 
much attention in the literature both for the study of theoretical 
properties and as a model for data fitting  in empirical work.
A major motivation for this interest is the high degree of flexibility of 
the distribution as the parameters span their admissible range, 
with ample variation of the associated measures of skewness and kurtosis.
While this high flexibility allows to adapt a member of the parametric
family to a wide range of data patterns, it also implies that parameter
estimation is a more delicate operation with respect to less flexible 
parametric families, given that a small variation of the parameters can 
have a substantial effect on the selected distribution.
In this context, the aim of the present contribution is to deal with 
some computational aspects of maximum likelihood estimation. 
A problem of interest is the possible presence of multiple local maxima
of the log-likelihood function. 
Another one, to which most of our attention is dedicated, is the development 
of a quick and reliable initialization method for the subsequent numerical 
maximization of the log-likelihood function, 
both in the univariate and the multivariate context. }

%------------------------------------------------------------------------------

\section{Background and aims}
\label{s:intro}
% large data-sets, flexible distributions, connection to robustness (?)
 
\subsection{Flexible distributions: the skew-$t$ case} \label{s:flexi-ST}

In the context of distribution theory, a central theme is the study of 
flexible parametric families of probability distributions, that is, 
families allowing substantial variation of their behaviour when the
parameters span their admissible range. For brevity, we shall refer
to this domain with the phrase `flexible distributions'.
The archetypal construction of this logic is represented by the Pearson 
system of curves for univariate continuous variables. 
In this formulation, the density function is regulated by four parameters,
allowing wide variation of the measures of skewness and kurtosis, hence
providing much more flexibility than in the basic case represented by the 
normal distribution, where only location and scale can be adjusted.

Since Pearson times, flexible distributions have remained a persistent
theme of interest in the literature, with a particularly intense 
activity in recent years. 
A prominent feature of newer developments is the increased consideration 
for multivariate distributions, reflecting the current availability in
applied work of larger datasets, both in sample size and in dimensionality.
In the multivariate setting, the various formulations often feature 
four blocks of parameters to regulate location, scale, skewness and kurtosis.

While providing powerful tools for data fitting, flexible distributions
also pose some challenges when we enter the concrete estimation stage. 
We shall be working with maximum likelihood estimation (MLE) 
or variants of it, 
but qualitatively similar issues exist for other criteria.
Explicit expressions of the estimates are out of the question; 
some numerical optimization procedure is always involved and 
this process is not so trivial because of the larger number of
parameters involved, as compared with fitting simpler parametric 
models, such as a Gamma  or a Beta distribution.
Furthermore, in some circumstances, the very flexibility of 
these parametric families can lead to difficulties: 
if the data pattern does not aim steadily towards 
a certain point of the parameter space, there could be two or more 
such points which constitute comparably valid candidates in terms
of log-likelihood or some other estimation criterion.
Clearly, these problems are more challenging with small sample size, 
later denoted $n$, since the log-likelihood function (possibly tuned by a prior 
distribution) is relatively more flat, but numerical experience has 
shown that they can persist even for fairly large $n$, in certain cases.

The focus of interest in this paper will be the skew-$t$ (ST) 
distribution  introduced by \cite{bran:dey:2001} and  studied 
in detail by \cite{azza:capi:2003}; see also \cite{guptaAK:2003}. 
The main formal constituents and properties of the ST family will
be summarized in the next subsection.
Here, we recall instead some of  the many publications that have  
provided evidence of the practical usefulness of the ST
family, in its univariate and multivariate version, thanks to
its capability to adapt to a variety of data patterns.
The numerical exploration by \cite{azza:gent:2008}, using data 
of various origins and nature, is an early  study  in this direction, 
emphasizing the potential of the distribution as a tool for robust inference. 
The robustness aspects of ST-based inference has also been discussed
by \cite[\S\,4.3.5]{azza:capi:2014} and more extensively 
by \cite{azzalini:2016icors}.
On the more applied domain, numerous publications motivated by 
application problems have further highlighted the ST usefulness,
typically with data distributions featuring substantial tailweight
and asymmetry.
For space reasons, the following list reports only a few of the many 
publications of this sort, with a preference for early work:
\cite{walls:2005} and \cite{pitt:2010jce}   
use the ST distribution  for modelling log-transformed returns of film 
and music industry products as a function of explanatory variables;
\cite{meucci:2006a}  and \cite{adcock:2010}  develop methods 
for optimal portfolio allocation in a financial context, where
long tails and asymmetry of returns distribution are standard features;
\cite{ghiz:roth:ruda:2010} use the multivariate ST distributions to 
model riverflow, jointly at multiple sites; 
%\cite{padoan:2011} 
% Multivariate extreme models based on underlying skew-t and skew-normal distributions
\cite{pyne:huX:etal:2009pnas} present an early model-based clustering 
formulation using the multivariate ST distributions as the basic component
for flow cytometric data analysis.

Given its value in data analysis, but also the above-mentioned possible 
critical aspects of the log-likelihood function, it seems appropriate to
explore the corresponding issues for MLE computation and to develop
a methodology  which provides good starting points for the
numerical maximization of the log-likelihood. 
After a brief summary of the main facts about the ST distribution in
the next subsection, the rest of the paper is dedicated to these issues.
Specifically, one section examines qualitatively and numerically various 
aspects of the ST log-likelihood, while the rest of the paper develops a
technique to initialize the numerical search for MLE.

To avoid potential misunderstanding, we underline that the above-indicated 
program of work does not intend to imply a general inadequacy of the currently 
available computational resources, which will be recalled in due course. 
There are, however, critical cases where these resources run into problems, 
most typically when the data distribution exhibits very long tails.
For these challenging situations, an improved methodology is called for.
 
%------------------------------------------------------------------------------
\subsection{The skew-$t$ distribution: basic facts} \label{s:ST-facts}

Before entering our actual development, we recall some basic facts
about the  ST parametric family of continuous distributions.
In its simplest description, it is obtained as a perturbation of
the classical Student's $t$ distribution.
For a more specific description, start from the univariate setting, 
where the components of the family are identified by four parameters. 
Of these four  parameters,  the one denoted $\xi$ in the following
regulates the location of the distribution;
scale is regulated by the positive parameter $\omega$; 
shape (as departure from symmetry) is regulated by $\lambda$;
tail-weight is regulated by $\nu$ (with $\nu>0$), 
denoted `degrees of freedom' like for a classical $t$ distribution.

It is convenient to introduce the distribution  in the `standard case', 
that is, with location $\xi=0$ and scale $\omega=1$.
In this case, the density function is
\begin{equation}
  t(z; \lambda, \nu) = 
  2\:t(z;\nu)\:T\left(\lambda z \sqrt{\frac{\nu+1}{\nu+z^2}}; \nu+1\right),
  \qquad z\in\Real,
       \label{e:st-pdf}
\end{equation}
where 
\begin{equation} 
  t(z;\nu) = \frac{\Gamma\big(\half(\nu+1)\big)}%
                  {\sqrt{\pi\,\nu}\:\Gamma\big(\half\nu\big)}
              \:\left(1+\frac{z^2}{\nu}\right)^{-(\nu+1)/2}, 
   \qquad z\in\Real,
   \label{e:t-pdf}
\end{equation}  
is the density function of the classical Student's $t$ on $\nu$ degrees
of freedom and $T(\cdot;\nu)$ denotes its distribution function;
note however that in (\ref{e:st-pdf}) this is evaluated with $\nu+1$ 
degrees of freedom.  Also, note that the symbol $t$ is used for both
densities in  (\ref{e:st-pdf}) and (\ref{e:t-pdf}), which are distinguished
by the presence of either one or two parameters.

If $Z$ is a random variable with density function (\ref{e:st-pdf}), 
the location and scale transform $Y=\xi+\omega\,Z$ has density  
\begin{equation}
     t_Y(x; \theta) = \omega\inv\: t(z; \lambda, \nu), \qquad 
     z= \omega\inv(z-\xi),
     \label{e:st-pdf-4par}
\end{equation}
where $\theta=(\xi,\omega,\lambda,\nu)$. 
In this case, we write $Y \sim \ST(\xi,\omega^2,\lambda,\nu)$, where $\omega$ 
is squared for similarity with the usual notation for normal distributions. 

When $\lambda=0$, we recover the scale-and-location family generated by
the $t$ distribution (\ref{e:t-pdf}). When $\nu\to\infty$, we obtain the
skew-normal (SN) distribution with parameters $(\xi,\omega,\lambda)$, 
which is described for instance by \cite[Chap.\,2]{azza:capi:2014}.
When $\lambda=0$ and $\nu\to\infty$, (\ref{e:st-pdf-4par}) converges
to the $\mathrm{N}(\xi, \omega^2)$ distribution.

\begin{figure}% [ht]
\includegraphics[width=0.49\textwidth]{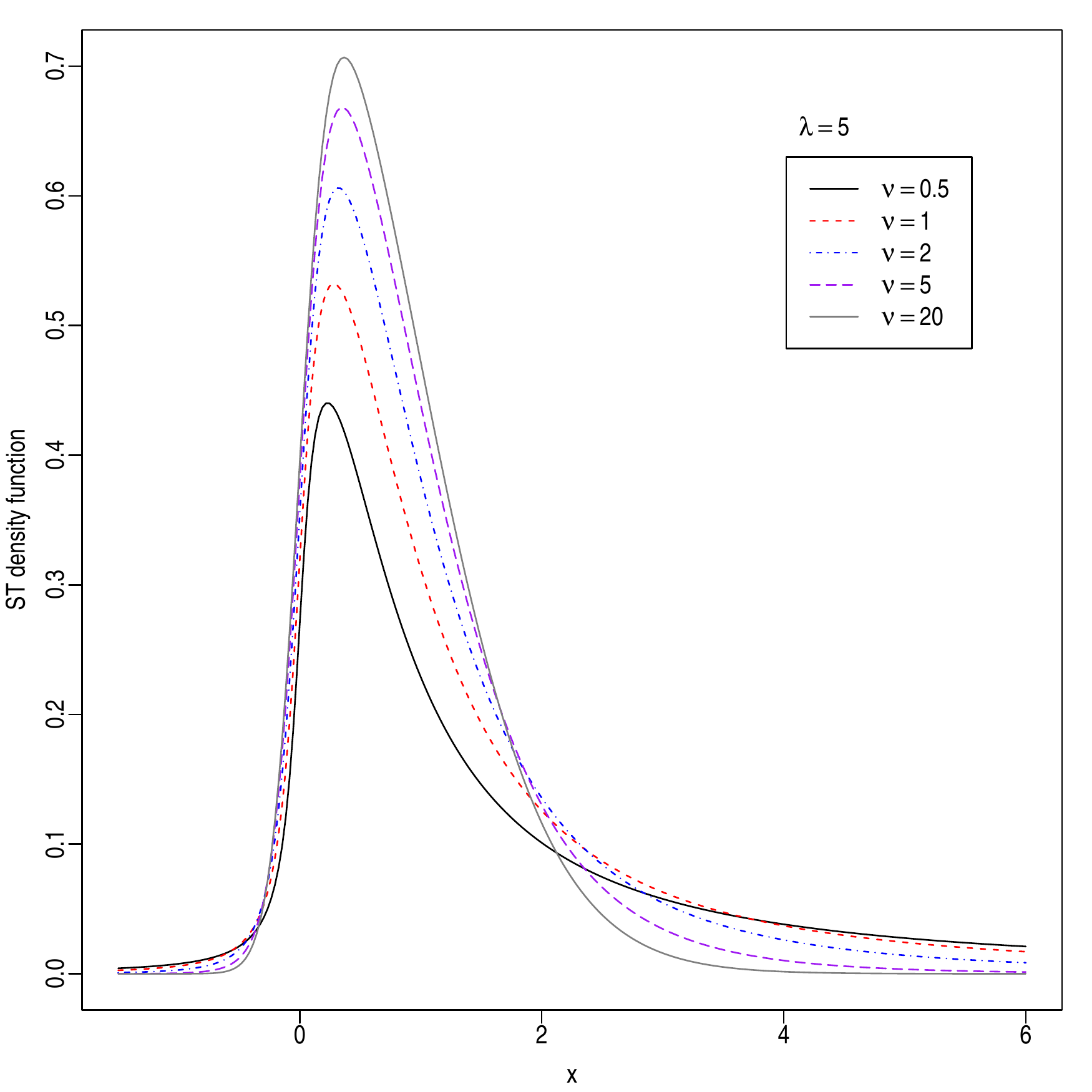}\quad
\includegraphics[width=0.49\textwidth]{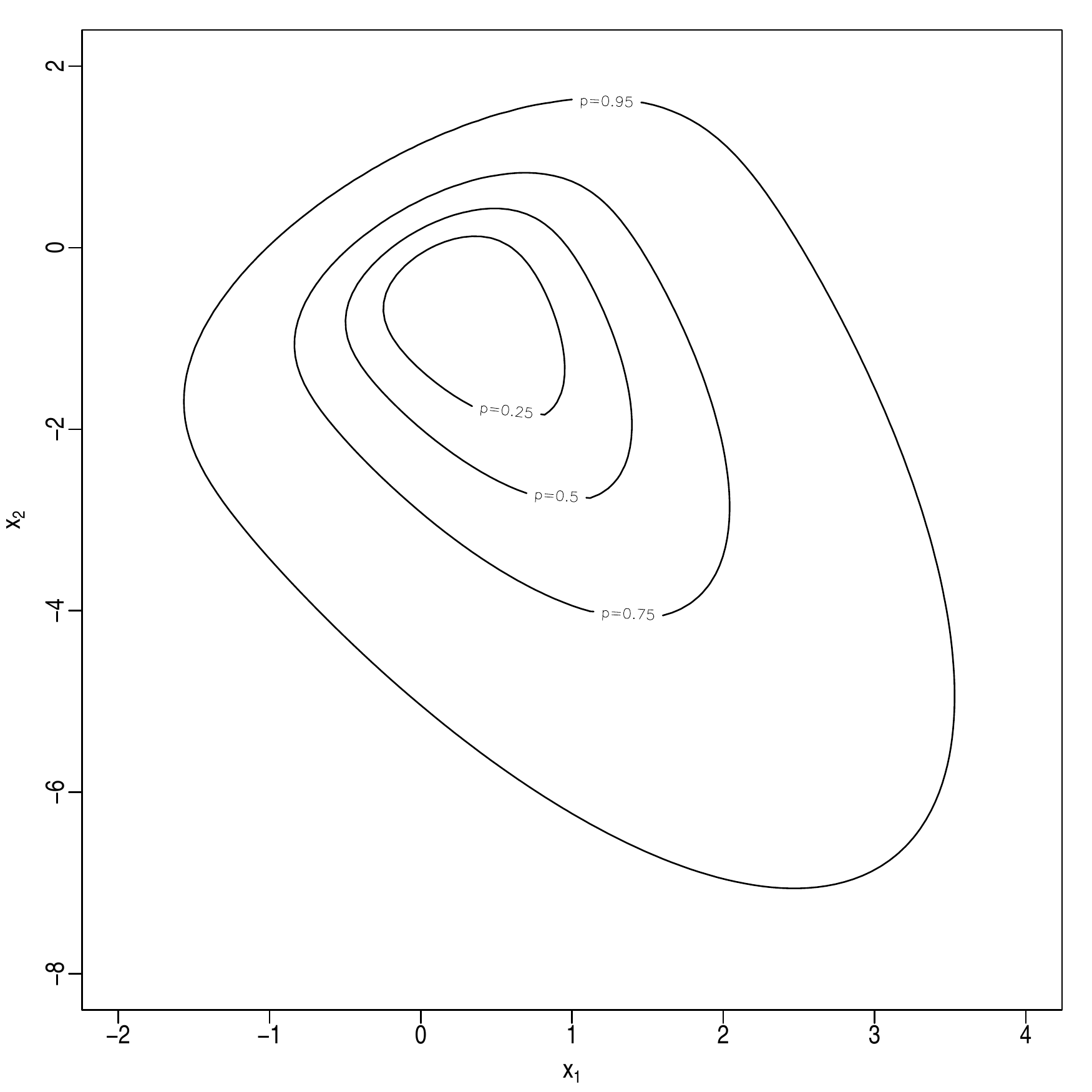}
\caption{The left plot displays a set of univariate skew-$t$ density functions 
   when $\lambda=5$  and $\nu$ varying across a range of values; 
   the right plot displays the contour level plot of a bivariate skew-$t$ density.}
\label{f:pdf}
\end{figure}

Some instances of density (\ref{e:st-pdf}) are displayed in the left pane
of Figure~\ref{f:pdf}. If $\lambda$ was replaced by $-\lambda$,
the densities would be reflected on the opposite side 
of the vertical axis, since $-Y\sim\ST(-\xi, \omega^2,-\lambda, \nu)$.

Similarly to the classical $t$ distribution, moments exist when 
their order is smaller than $\nu$.  Under this condition,
expressions for the mean value, the variance and the coefficients of skewness
and excess kurtosis, namely the standardized third and fourth cumulants, are 
as follows:
\begin{eqnarray*}
  \mu &=& \E{Y} 
     =\xi+\omega\, b_\nu\, \delta, \hspace{12.2em} \hbox{if~}\nu>1, 
           \label{e:st-mean} \\
  \sigma^2 &=& \mathrm{var}\{Y\} 
     = \omega^2\left[\frac{\nu}{\nu-2}- (b_\nu\:\delta)^2\right]
         =\omega^2\sigma^2\subZ\,, \hbox{~say,} \hspace{2em}\hbox{if~}\nu>2,
            \label{e:st-var}  \\
  \gamma_1 &=&   \frac{b_\nu\,\delta}{\sigma\subZ^{3}}\,\,
   \left[\frac{\nu(3-\delta^2)}{\nu-3}
         - \frac{3\,\nu}{\nu-2}+2\,(b_\nu\,\delta)^2\right]\,,
   % \left[\frac{\nu}{\nu-2}-(b_\nu\,\delta)^2\right]^{-3/2}\,
   \hspace{4.2em} \hbox{if~}\nu>3,   \label{e:st-gamma1} \\
  \gamma_2 &=& \frac{1}{\sigma\subZ^4}\,\left[  \frac{3\nu^2}{(\nu-2)(\nu-4)}
      - \frac{4(b_\nu\,\delta)^2\nu(3-\delta^2)}{\nu-3}
      + \frac{6(b_\nu\,\delta)^2\nu}{\nu-2}
      - 3(b_\nu\,\delta)^4\right] - 3\,           \nonumber\\
     % \left[\frac{\nu}{\nu-2}-(b_\nu\,\delta)^2\right]^{-2} -3, \nonumber\\
     && \hspace{20em}     \hbox{~if~}\nu>4,  \label{e:st-gamma2} 
\end{eqnarray*}  
where
\begin{equation}
  \delta = \delta(\lambda)
         = \frac{\lambda}{(1+\lambda^2)^{1/2}}\in(-1,1),\hspace{2em}
  b_\nu = \frac{\sqrt{\nu}\: \Gamma\left(\half(\nu-1)\right)}
           {\sqrt{\pi}\:\Gamma\left(\half\nu\right) }
           \quad \hbox{if~}\nu>1.
  \label{e:delta,b.nu}           
\end{equation}

It is visible that, as $\lambda$ spans the real line, so does the 
coefficient of skewness $\gamma_1$ when $\nu\to3^+$. 
For $\nu\le3$,  $\gamma_1$ does not exists; 
however, at least one of the tails is increasingly heavier
as $\nu\to0$, given the connection with the Student's $t$.
A fairly similar pattern holds for the coefficients of kurtosis $\gamma_2$, 
with the threshold  at $\nu=4$ for its existence. 
If $\nu\to4^+$, the range of $\gamma_2$ is $[0,\infty$).
The feasible $(\gamma_1, \gamma_2)$ space when $\nu>4$ is displayed
in Figure~4.5 of \cite{azza:capi:2014}.
Negative $\gamma_2$ values are not achievable, 
but this does not not seem to be a major drawback in most applications.

The multivariate ST density is represented by a  perturbation of the 
classical multivariate $t$ density in $d$ dimensions, namely
\begin{equation} \label{e:mt-pdf}
  t_d(z; \bar\Omega,\nu) =  \frac{\Gamma((\nu+d)/2)}%
        {(\nu\pi)^{d/2} \,\Gamma(\nu/2)\,\det(\bar\Omega)^{1/2}}
        \left(1+\frac{Q(z)}{\nu}\right)^{-\frac{\nu+d}{2}}  \,,
        \quad z\in\Real^d \,.
\end{equation}
where $\bar\Omega$ is a symmetric positive-definite matrix with unit 
diagonal elements and $Q(z)=z\T\bar\Omega\inv z$.
The multivariate version of (\ref{e:st-pdf})  is then given by
\begin{equation} \label{e:mst-pdf} 
  t_d(x) =  2\: t_d(z;\bar\Omega,\nu)\:    
            T\left(\alpha\T z\sqrt{\frac{\nu+d}{\nu+Q(z)}}; \nu+d \right)\,,
            \qquad  z\in\Real^d 
\end{equation}  
where $\alpha$ is a $d$-dimensional vector regulating asymmetry. 
An instance of density (\ref{e:mst-pdf}) with $d=2$ is displayed 
in the right pane of Figure~\ref{f:pdf} via contour level curves.

Similarly to the univariate setting, we consider the location and scale
transformation of a variable $Z$ with density (\ref{e:mst-pdf})
to $Y=\xi+\omega\,Z$ where now $\xi\in\Real^d$ and $\omega$ is diagonal
matrix with positive diagonal elements.
For the resulting variable, we use the notation
$Y\sim \ST_d(\xi, \Omega, \alpha, \nu)$ where $\Omega=\omega\bar\Omega\omega$. 

One property required for our later development is that each marginal 
component of $Y$ is a univariate ST variable, having a density of type
(\ref{e:st-pdf}) whose parameters are extracted  from the corresponding
components of $Y$, with the exception of $\lambda$ for which the
marginalization step is slightly more elaborate. 
%Specifically, from  
%\begin{equation}
%   \delta = (\delta_1,\dots \delta_d)\T
%        =  (1+ \alpha\T\bar\Omega\alpha)^{-1/2}\bar\Omega\alpha\,,
%  \label{e:alpha->delta}    
%\end{equation}
%the vector of shape parameters $\lambda=(\lambda_1,\dots,\lambda_d)\T$ is
%obtained by component-wise inversion of  the first expression in 
%(\ref{e:delta,b.nu}), namely as 
%\[
%   \lambda_j = (1-\delta_j^2)^{-1/2}\,\delta_j \,,\qquad j=1,\dots,d\,.
%\]

There are many additional properties of the ST distribution which, for
space reasons, we do not report here and refer the reader to the quoted
literature. 
A self-contained account is provided by the monograph \cite{azza:capi:2014};
see specifically Chapter~4 for the univariate case and Chapter~6 for 
the multivariate case.

%------------------------------------------------------------------------------
\section{On the likelihood function of ST models}  \label{s:ST-logL}

%\begin{obeylines}\tt
%Components of this section:
%Recall basic facts about logL (no singularity, etc.)
%Estimates on the boundary and MPLE.
%Mention the problem of poles (see \cite{azza:capi:2003})
%Software tools ('sn', optim and nlminb)
%A selection of critical cases/examples of logL behaviour.
%Estimation of the probability of multiple maxima.
%\end{obeylines}
%\vspace{2ex}

\subsection{Basic general aspects}

The high flexibility of the ST distribution makes it particularly appealing 
in a wide range of data fitting problems, more than its companion, 
the SN distribution. Reliable techniques for implementing
connected MLE or other estimation methods are therefore crucial.

From the inference viewpoint, another advantage of the ST over 
the related SN distribution is the lack of a stationary point
at $\lambda=0$ (or $\alpha=0$ in the multivariate case), and the 
implied singularity of the information matrix. 
This stationary point of the SN is systematic: 
it occurs for all samples, no matter what $n$ is.
This peculiar aspect has been emphasized more than necessary 
in the literature,  considering that it pertains to a single although 
important value  of the parameter. 
Anyway, no such problem exists under the ST assumption. 
The lack of a stationary point at the origin was first observed
empirically and welcomed as `a pleasant surprise' by \cite{azza:capi:2003},
but no theoretical explanation was given.
Additional numerical evidence in this direction has been provided by
\cite{azza:gent:2008}. 
The theoretical explanation of why the SN and the ST likelihood functions 
behave differently was finally established by \cite{hall:ley:2012}.

Another peculiar aspect of the SN likelihood function is the possibility
that the maximum of the likelihood function occurs at $\lambda=\pm\infty$,
or at $\|\alpha\|\to\infty$ in the multivariate case.
Note that this happen without divergence of the likelihood function,
but only with divergence of the parameter achieving the maximum.
In this respect the SN and the ST model are similar: 
both of them can lead to this pattern.

Differently from the stationarity point at the origin, the phenomenon of 
divergent estimates is transient: it occurs mostly with small $n$,
and the probability of its occurrence decreases very rapidly 
when $n$ increases. 
However, when it occurs for the $n$ available data, we must handle it.
There are different views among statisticians on whether such divergent 
values must be retained as valid estimates or they must be rejected 
as unacceptable. 
We embrace the latter view, for the reasons put forward by 
\cite{azza:arel:2013jspi}, and adopt the maximum penalized likelihood
estimate (MPLE) proposed there to prevent the problem. 
While the motivation for MPLE is primarily for small to moderate  $n$,
we use it throughout for consistency. 
% The MPLE methodology is implemented in the \R\ package \texttt{sn}; 
% see \cite{azzalini:sn1.5-4}. 

% Specifically, we shall make use of functions \texttt{st.mle} 
%and \texttt{mst.mple}
  
There is an additional peculiar feature of the ST log-likelihood 
function, which however we mention only for completeness, 
rather than for its real relevance. 
In cases when $\nu$ is allowed to span the whole positive half-line,
poles of the likelihood function must exist near $\nu=0$,
similarly to the case of a Student's $t$ with unspecified
degrees of freedom.
This problem has been explored numerically by \cite{azza:capi:2003},
and the indication was that these poles must exist at \emph{very}
small values of $\nu$, such as $\hat\nu=0.06$ in one specific instance.

This phenomenon is qualitatively similar to the problem of poles of the 
likelihood function for a finite mixture of continuous distributions.
Even in the simple  case of univariate normal components, there always 
exist $n$ poles on the boundary of the parameter space 
if the standard deviations of the components are unrestricted; 
see for instance \cite[Section 7]{day:1969}.
%  Penalized Maximum Likelihood Estimator for Normal Mixtures
% GABRIELA CIUPERCA  ANDREA RIDOLFI  JÉRÔME IDIER
% First published: 28 February 2003 
% https://doi.org/10.1111/1467-9469.00317 Cited by: 46
The problem is conceptually interesting, in both settings, 
but in practice it is easily dealt with in various ways. 
In the ST setting, the simplest solution is to impose a constraint 
$\nu>\nu_0>0$ where $\nu_0$ is some  very small value, 
such as $\nu_0=0.1$ or $0.2$. 
Even if fitted to data, a $t$ or ST density with $\nu<0.1$ 
would be an object hard to use in practice.

%------------------------------------------------------------------------------
\subsection{Numerical aspects and some illustrations} \label{s:examples}

%\textsl{A selection of critical cases/examples of logL behaviour.}\\
%\indent\textsl{Estimation of the probability of multiple maxima.}
%\vspace{2ex}

Since, on the computational side, we shall base our work the \R\ package 
\texttt{sn}, described by \cite{azzalini:sn1.5-4},
it is appropriate to describe some key aspects of this package..
There exists a comprehensive function for model fitting, called \texttt{selm},
but the actual numerical work in case of an ST model is performed by
functions \texttt{st.mple} and \texttt{mst.mple}, in the univariate and
the multivariate case, respectively.
To numerical efficiency, we shall be using these functions directly,
rather than via \texttt{selm}.
As their names suggest, \texttt{st.mple} and \texttt{mst.mple} perform MPLE, 
but they can be used for classical MLE as well, 
just by omitting the penalty function.
The rest of the description refers to \texttt{st.mple}, but  
 \texttt{mst.mple} follows a similar scheme.
 
In the univariate case, denote by $\theta=(\xi, \omega, \alpha, \nu)\T$ 
the parameters to be estimated, or possibly 
$\theta=(\beta\T, \omega, \alpha, \nu)\T$ when
a linear regression model is introduced for the location parameter,
in which case $\beta$ is a vector of  $p$ regression coefficients.
Denote by $\log L(\theta)$ the log-likelihood function at point $\theta$.
If no starting values are supplied, the  first operation of \texttt{st.mple} 
is to fit a linear model  to the available explanatory variables; 
this reduces to the constant covariate value 1 if $p=1$.
For the residuals from this linear fit, sample cumulants of order up 
to four are computed, hence including the sample variance. 
An inversion from these values to $\theta$ may or may not be
possible, depending on whether the third and fourth sample cumulants
fall in the feasible region for the ST family. If the inversion is 
successful, initial values of the parameters are so obtained; if not, 
the final two components of $\theta$ are set at $(\alpha, \nu)=(0, 10)$, 
retaining the other components from the linear fit.
Starting from this point, MLE or MPLE is searched for using
a general numerical optimization procedure.
The default procedure for performing this step is the \R\ function 
\texttt{nlminb}, supplied with the score functions besides 
the log-likelihood function. 
We shall refer, comprehensively, to this currently standard procedure 
as `method M0'.

In all our numerical work, method M0 uses \texttt{st.mple}, and the involved 
function \texttt{nlminb}, with all tuning parameters kept at their default
values. The only activated option is the one switching between MPLE and
MLE, and even this only for the work of present section. 
Later on, we shall always use MPLE, with penalty function \texttt{Qpenalty}
which implements the method proposed in \cite{azza:arel:2013jspi}. 
 
We start our numerical work with some illustrations, essentially in graphical 
form, of the log-likelihood generated by some simulated datasets.
The aim is to provide a direct perception, although inevitably limited, 
of the possible behaviour of the log-likelihood and the ensuing problems 
which it poses for MLE search and other inferential procedures.
Given this aim, we focus on cases which are unusual, in some way or another,
rather than on `plain cases'.

The type of graphical display which we adopt is based on the profile 
log-likelihood function of $(\alpha, \nu)$, denoted $\log L_p(\alpha, \nu)$.
This is obtained, for any given $(\alpha, \nu)$, by maximizing
$\log L(\theta)$ with respect  to the remaining parameters.
To simplify readability, we transform  $\log L_p(\alpha, \nu)$ to the
likelihood ratio test statistic, also called `deviance function':
\begin{equation}
  D(\alpha, \nu)= 2\:\{\log L_p(\hat\alpha, \hat\nu)- \log L_p(\alpha, \nu)\}
  \label{e:deviance}
\end{equation}
where $\log L_p(\hat\alpha, \hat\nu)$ is the overall maximum value of 
the log-likelihood, equivalent to $\log L(\hat\theta)$.
The concept of deviance applies equally to the penalized log-likelihood.

%This means that $D(\alpha,\nu)\ge 0$ and the contour level curves of the 
%function at levels given by the $\chi^2_2$ quantiles can be interpreted
%as approximate confidence regions.

\begin{figure}% [ht]
\includegraphics[width=0.49\textwidth]{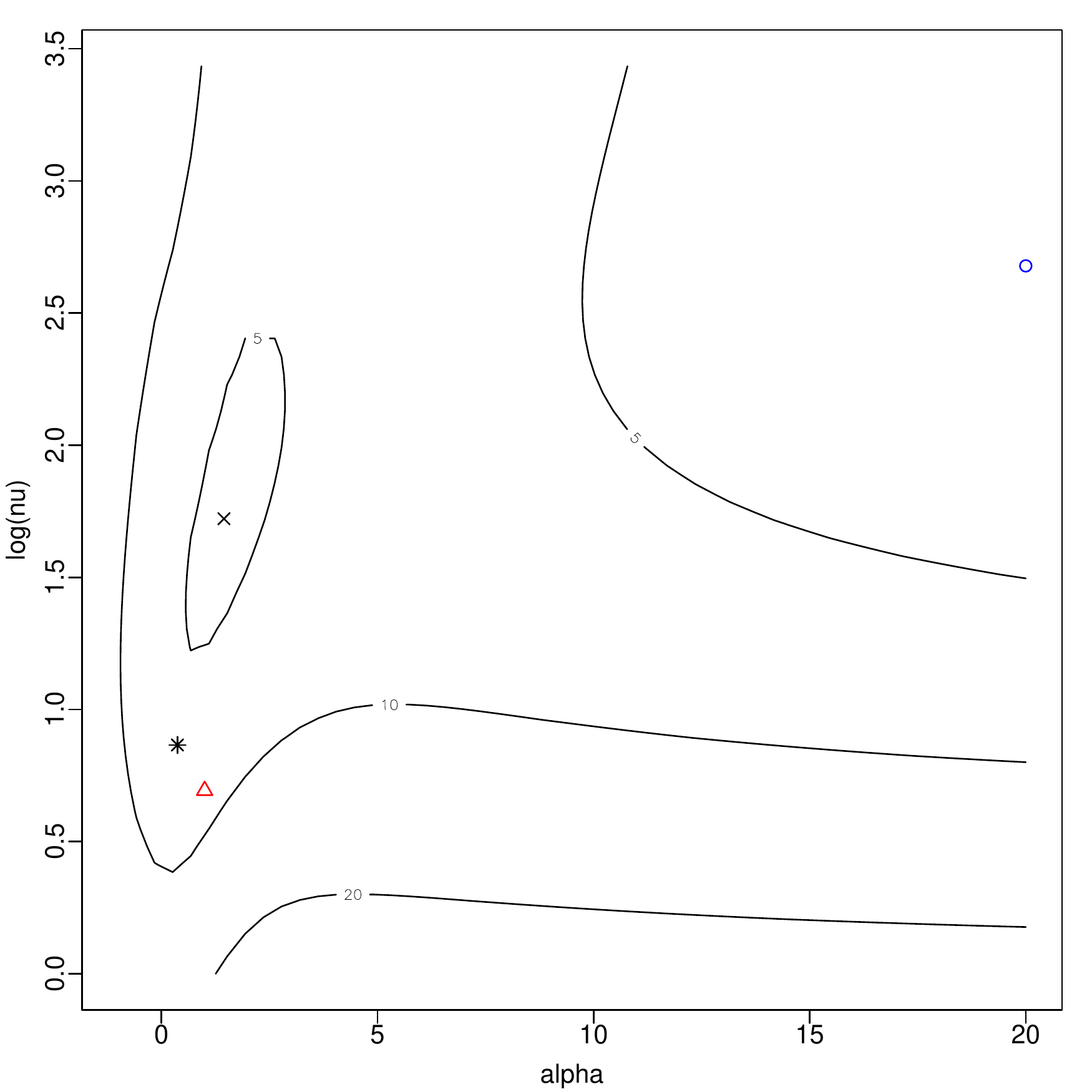}\quad
\includegraphics[width=0.49\textwidth]{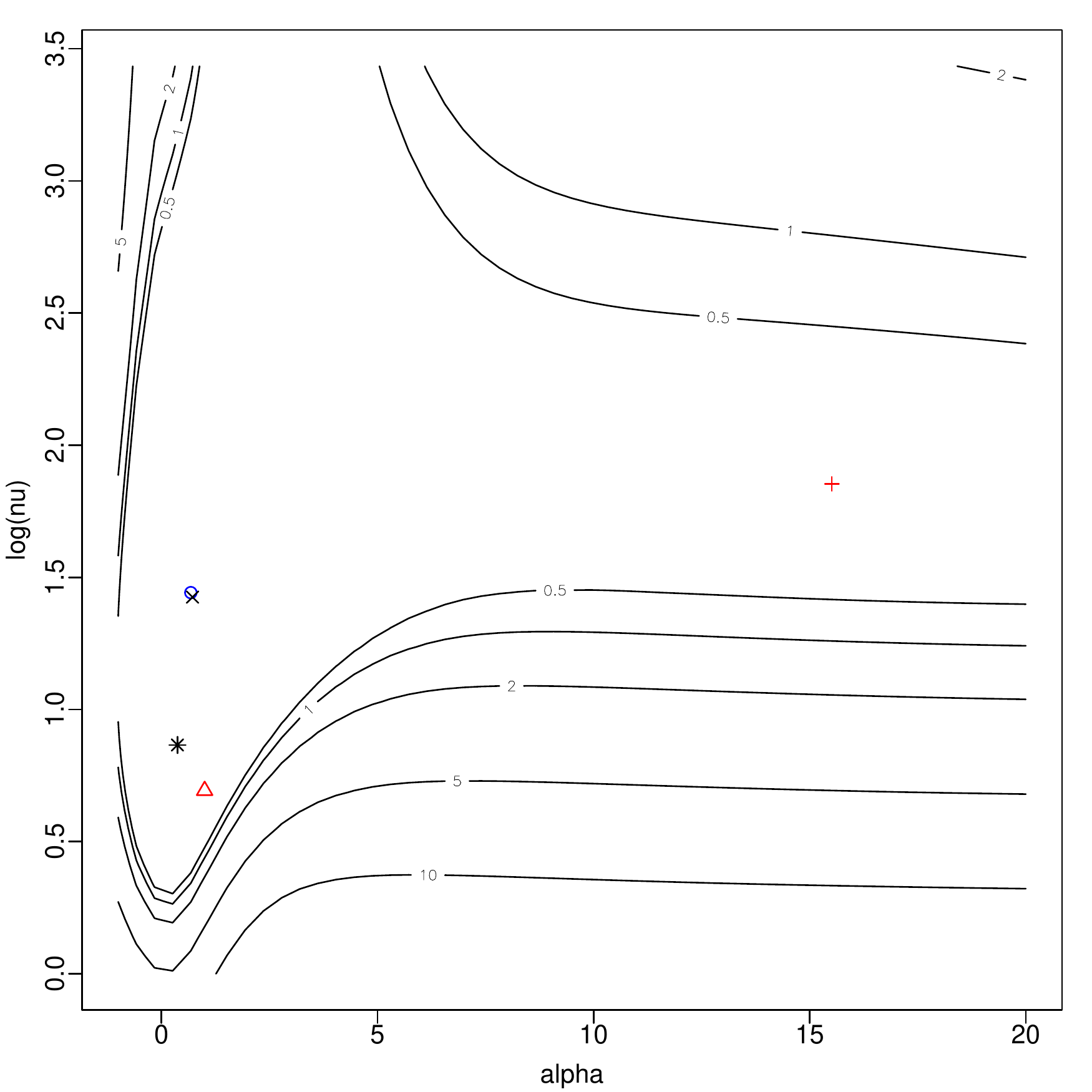}
\par
\includegraphics[width=0.49\textwidth]{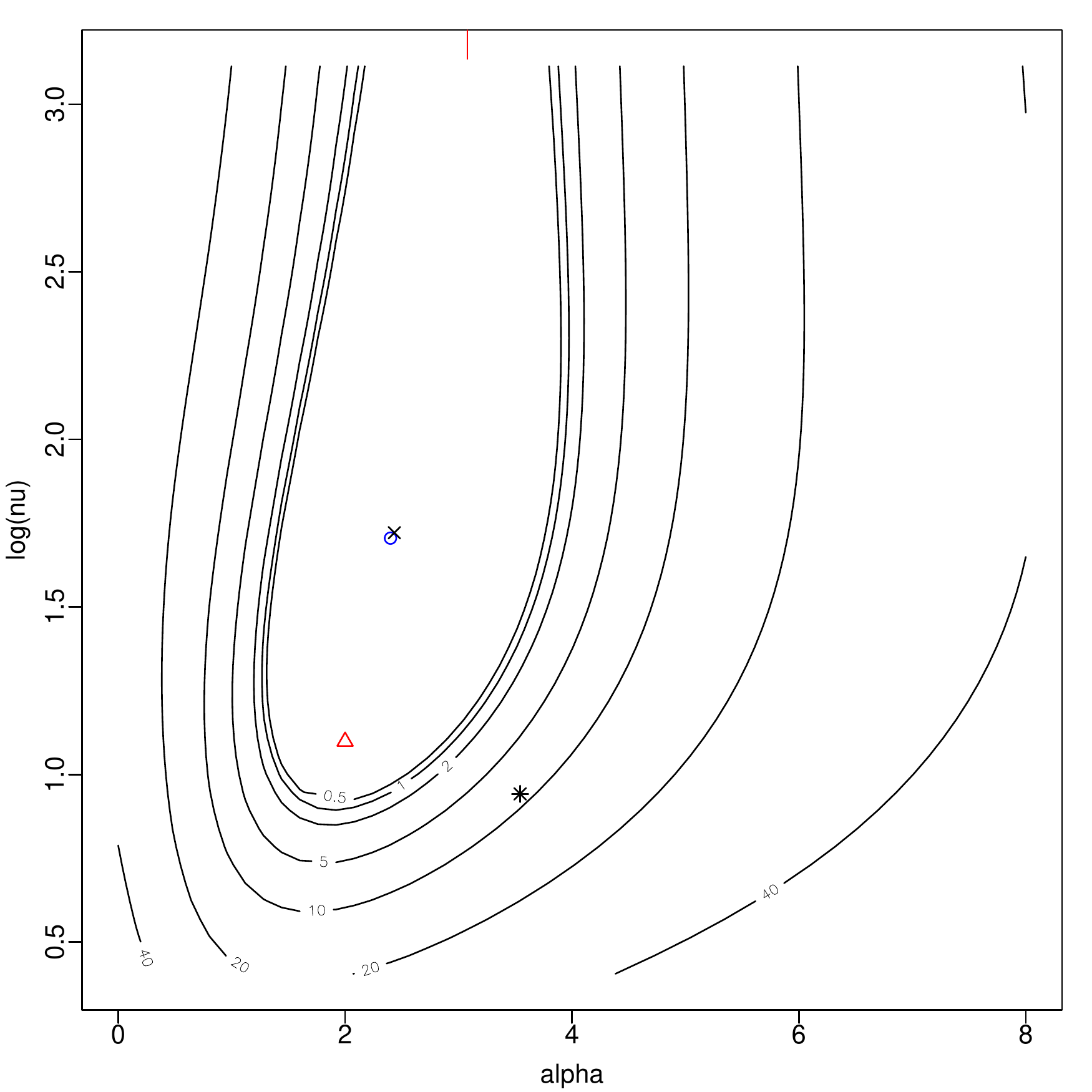}\quad
\includegraphics[width=0.49\textwidth]{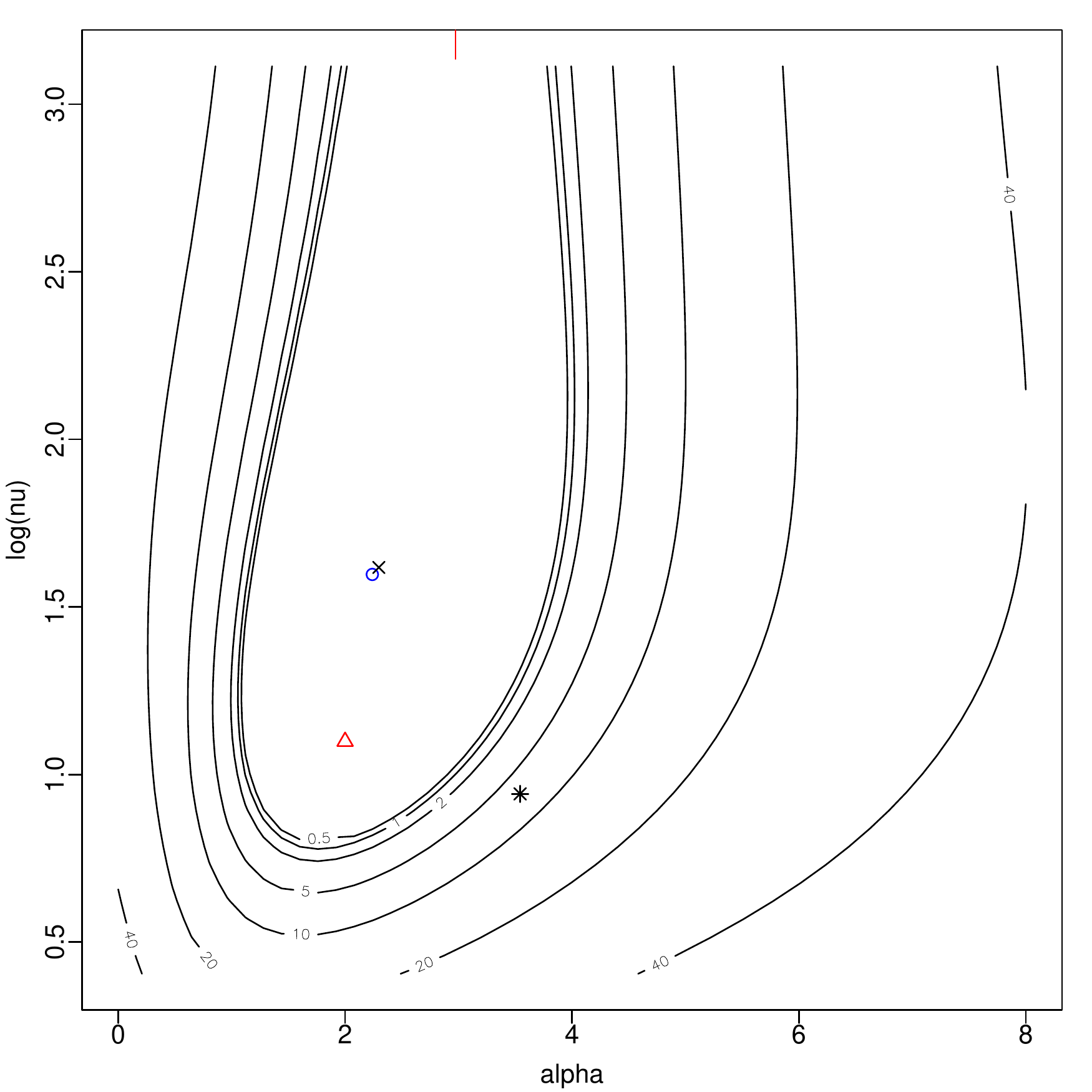}
\caption{Contour level plots of the 
 deviance function for simple samples from $\ST(0,1,\alpha, \nu)$
 associated to the log-likelihood (on the left side) or 
 to its penalized version (on the right side). 
 The top plots refer to a sample of size $n=50$ from $\ST(0,1, 1, 2)$;
 the bottom plots refer to a sample of size $n=300$ from $\ST(0,1, 2, 3)$.
 See the text for explanation of the marked points.}
\label{f:deviance}
\end{figure}

The plots in Figure~\ref{f:deviance} displays, in the form of contour
level plots, the behaviour of $D(\alpha, \nu)$ for two artificially 
generated samples, with $\nu$ expressed on the logarithmic scale 
for more convenient readability.
Specifically,  the top plots refer to a sample of size $n=50$ drawn 
from the $\ST(0,1, 1, 2)$;  
the left plot, refers to the regular log-likelihood, 
while the right plot refers to the penalized log-likelihood.
%-----------------
%-----------------
The plots include marks for points of special interest, as follows:
\begin{itemize}
\item[{\color{red}$\triangle$}] the true parameter point;
\item[{\large\color{blue}$\circ$}] the point having maximal (penalized) 
    log-likelihood on a $51\times51$ grid of points spanning the plotted area;
\item[{\color{red}$+$}] the MLE or MPLE point selected by method M0;    
\item[{\large $\ast$}] the preliminary estimate to be introduced in 
    Section~\ref{s:prelimFit-core}, later denoted M1;    
\item[$\times$] the  MLE or MPLE point selected by method M2
   presented later in the text.   
\end{itemize}

It will be noticed that the top-left plot does not show a {\color{red}$+$} mark. 
This is because the MLE point delivered by M0 has $\hat\alpha=\hat\nu\to\infty$
(actually some huge values representing numerical `approximations
of infinity'), where  $\log L_p\approx -81.85$; consequently the maximum
of $\log L_p$ over the plotted area takes place at its margin.
Note that the log-likelihood function has a local maximum at about
$(1.45, 5.6)$, where $\log L_p\approx -84.09$;
this local maximum is quite close to the true parameter point, 
especially so in the light of the limited sample size.
There are two messages from this example: one is that the log-likelihood
may have more than one maximum; the other is that a local maximum can
provide a better choice than the global maximum, at least in this case.

Given that $\hat\alpha=\infty$, consider MPLE estimation in the
top-right plot. The maximum of $\log L_p$, marked by {\color{blue}$\circ$}, 
is now close to the point $(1.45, 5.6)$, but method M0 fails to find it,
and it picks up the point $(15.5, 6.4)$. This must be due to a poor choice 
of the initial point for numerical search, given that method
M2, which differs only for this initial point, lands on the correct point.

Peculiar behaviours, either of the log-likelihood or of the estimation 
procedures or both of them,
are certainly more frequent when $n$ is small or moderate, 
but problems can persist even for fairly
large $n$, as illustrated by the bottom two plots of 
Figure~\ref{f:deviance} which refer to a sample of size $n=300$ 
from $\ST(0,1, 2, 3)$. In this case, M0 yields $\hat\alpha\approx 3$
and $\hat\nu\to\infty$, denoted by the vertical ticks below the 
top side of the plotted area; the associated $\log L_p$ value is
$-422.5$ for the left plot,  $-424.4$ for the right plot.
Both these values are lower than the corresponding 
maximal $\log L_p$ values, $-419.8$ and $-420.8$. 
Again, better initial search points used by method M2 leads to
the correct global maxima, at about $(2.4, 5.6)$ and $(2.3, 5.0)$, 
respectively.
  
%------------------------------------------------------------------------------
\section{On the choice of initial parameters for MLE search} 
\label{s:prelimFit}

The aim of this section, which represents the main body of the paper, 
is to develop a methodology for improving the selection of  initial 
parameter values from where to start the MPLE search via some numerical
optimization technique,  which should hopefully achieve a higher 
maximum.

%------------------------------------------------------------------------------
\subsection{Preliminary remarks and the basic scheme}  
\label{s:prelimFit-intro}

We have seen in Section~\ref{s:ST-logL} the ST log-likelihood function 
can be problematic; it is then advisable to select carefully 
the starting point for the MLE search.  
While contrasting the risk of landing on a local maximum, 
a connected aspect of interest is to reduce the overall computing time.
Here are some preliminary considerations about the stated target.

Since these initial estimates will be refined by a subsequent step of
log-likelihood maximization, there is no point in aiming at a very
sophisticate method. 
In addition, we want to keep the involved computing header as light as possible. 
Therefore, we want a method which is simple and quick to compute; 
at the same time, it should be reasonably reliable, 
hopefully avoiding nonsensical outcomes.

Another consideration is that we cannot work with the methods of moments,
or some variant of it, as this would impose a condition $\nu>4$, 
bearing in mind the constraints recalled in Section~\ref{s:ST-facts}.
Since some of the most interesting applications of ST-based models
deal with very heavy tails, hence with low degrees of freedom, 
the condition $\nu>4$ would be unacceptable in many important applications.
The implication is that we have to work with quantiles and derived
quantities.

To ease exposition, we begin by presenting the logic in the basic case 
of independent observations from a common univariate distribution 
$\ST(\xi, \omega^2, \lambda, \nu)$.
The first step is to select suitable quantile-based measures of location, 
scale, asymmetry and tail-weight. 
The following list presents a set of reasonably choices; 
these measures can be equally referred to a probability distribution or 
to a sample, depending on the interpretation of the terms quantile, 
quartile and alike.
\begin{description}
\item[\sl Location] 
  The median is the obvious choice here; denote it by $q_2$,
  since it coincides with the second quartile. 
\item[\sl Scale] 
  A commonly used measure of scale is the semi-interquartile difference,
  also called  quartile deviation, that is
  \[      d_q = \half(q_3- q_1)  \]   
  where $q_j$ denotes the $j$th quartile; 
  see for instance \cite[vol.\,10, p.\,6743]{ESS:2006}.
\item[\sl Asymmetry] A classical non-parametric measure of asymmetry 
  is the so-called Bowley's measure 
\[
   G = \frac{(q_3-q_2) - (q_2-q_1)}{q_3-q_1} 
     = \frac{q_3- 2\,q_2 + q_1}{2\,d_q}\,; 
\] 
see \cite[vol.\,12, p.\,7771--3]{ESS:2006}.
Since the same quantity, up to an inessential difference, had previously
been used by Galton, some authors attribute to him its introduction.
We shall refer to $G$ as the Galton-Bowley measure. 
\item[\sl Kurtosis] 
A relatively more recent proposal is the Moors measure of kurtosis, 
presented in \cite{moors:1988},
\[
   M =\frac{(e_7-e_5) + (e_3-e_1)}{e_6-e_2}
\]
where $e_j$ denotes the $j$th octile, for $j=1,\dots,7$.
Clearly, $e_{2j}=q_j$ for $j=1,2,3$.
% see also \cite[vol.\,6, p.\.3906]{ESS:2006}.
\end{description}
A key property is that $d_q$ is independent of the location of 
the distribution, and $G$ and $M$ are independent of location and scale.

For any distribution $\ST(\xi, \omega^2, \lambda, \nu)$,
the values  of $Q=(q_2, d_q, G, M)$  are functions of the parameters
$\theta=(\xi, \omega, \lambda, \nu)$.
Given a set of observations $y=(y_1, \dots, y_n)$ drawn 
from $\ST(\xi, \omega^2, \lambda, \nu)$ under mutual independence 
condition, we compute sample values of 
$\tilde{Q}=(\tilde{q}_2, \tilde{d}_q, \tilde{G}, \tilde{M})$  
of $Q$ from the sample quantiles and then inversion of the functions 
connecting $\theta$ and $Q$ will yield estimates $\tilde\theta$ 
of the ST parameters. 
In essence, the logic is similar to the one underlying the 
method of moments, but with moments replaced by quantiles.

In the following subsection, we discuss how to numerically carry out 
the inversion from $Q$ to $\theta$. Next, we extend the procedure to settings 
which include explanatory variables and  multivariate observations.

%------------------------------------------------------------------------------
\subsection{Inversion of quantile-based measures to ST parameters} 
\label{s:prelimFit-core}

For the inversion of the parameter set $Q=(q_2, d_q, G, M)$ to 
$\theta=(\xi, \omega, \lambda, \nu)$, the first stage considers only 
the components $(G,M)$ which are to be mapped  to $(\lambda, \nu)$,
exploiting the invariance of $G$ and $M$ with respect to location and scale.
Hence, at this stage, we can work assuming that $\xi=0$ and $\omega=1$. 
% the target is to devise a rule  for mapping sample values  
% $(\tilde{G}, \tilde{M})$ into initial estimates $(\tilde\lambda, \tilde\nu)$.

Start by computing, for any given pair $(\lambda, \nu)$, the set of octiles 
$e_1, \dots, e_7$ of $\ST(0,1,\lambda,\nu)$, and from here the corresponding 
$(G, M)$ values.
Operationally, we have computed the ST quantiles using routine \texttt{qst} 
of package \texttt{sn}.
Only non-negative values of $\lambda$ need to be considered, because 
a reversal of the $\lambda$ sign simply reverses the sign of $G$,
while $M$ is unaffected, thanks to the mirroring property of the
ST quantiles when $\lambda$ is changed to $-\lambda$.

% [block comment remove]

Initially, our numerical exploration of the inversion process examined
the contour level plots of $G$ and $M$ as functions of $\lambda$ and $\nu$,
as this appeared to be the more natural approach.
Unfortunately, these plots turned out not to be useful, because of the 
lack of a sufficiently regular pattern of the contour curves.
Therefore these plots are not even displayed here.

A more useful display is the one adopted in Figure~\ref{f:(G,M)space-sections}, 
where the coordinate axes are now $G$ and $M$. 
The shaded area, which is the same in both panels, represents the set 
of feasible $(G,M)$ points for the ST family.
In the first plot, each of the black lines indicates the \emph{locus} 
of points with constant values of $\delta$,  defined by (\ref{e:delta,b.nu}), 
when $\nu$ spans the positive half-line;
the selected $\delta$ values are printed at the top of the shaded area,
when feasible without clutter of the labels.
The use of $\delta$ instead of $\lambda$ simply yields a better spread
of the contour lines with different parameter values, but it is conceptually
irrelevant. 
The second plot of Figure~\ref{f:(G,M)space-sections} displays
the same admissible region with superimposed a different type of \emph{loci}, 
namely those corresponding to specified values of $\nu$, 
when $\delta$ spans  the $[0,1]$  interval;
the selected $\nu$ values are printed on the left side of the shaded area.

\begin{figure}
\centering
\includegraphics[width=0.49\hsize]{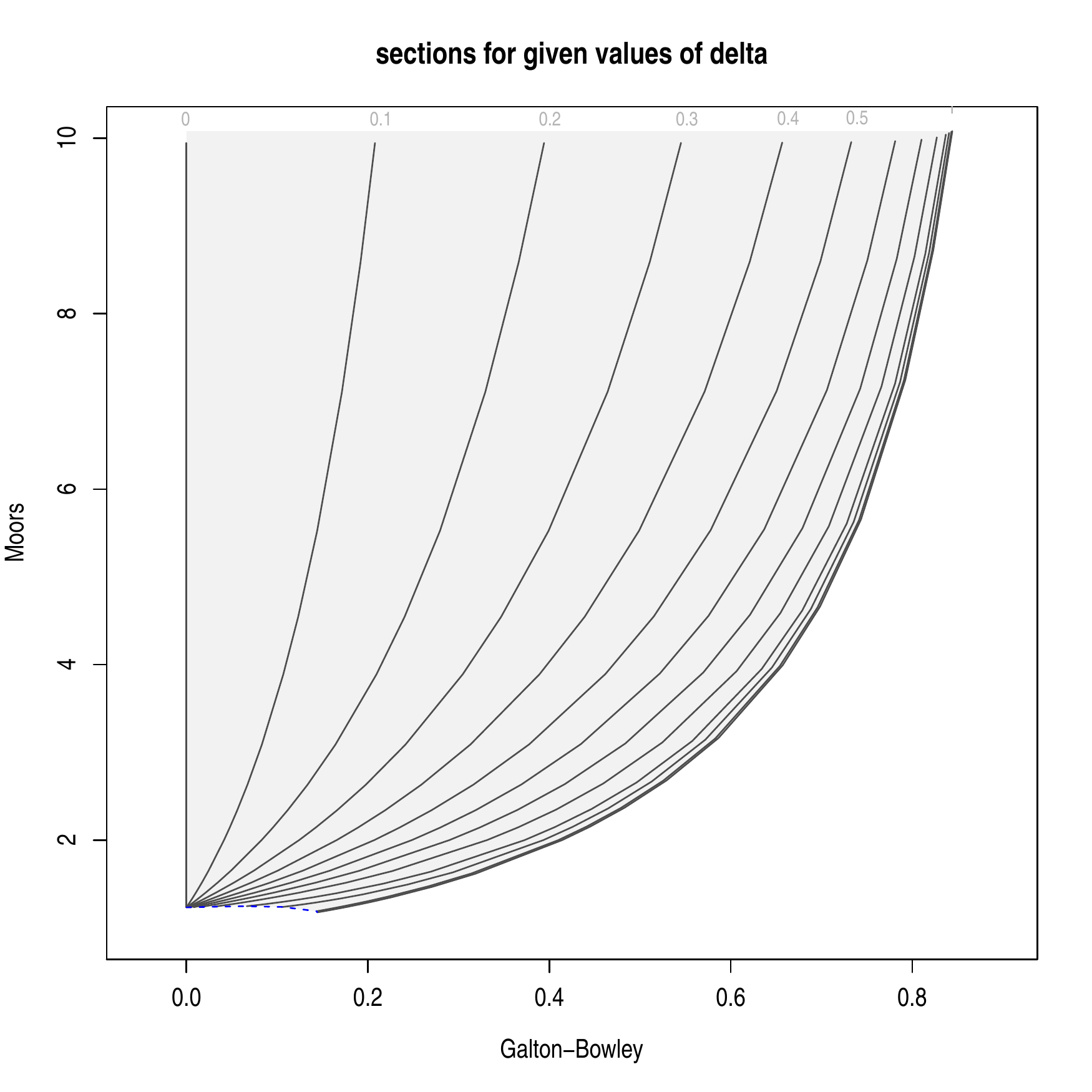}
\hfill
\includegraphics[width=0.49\hsize]{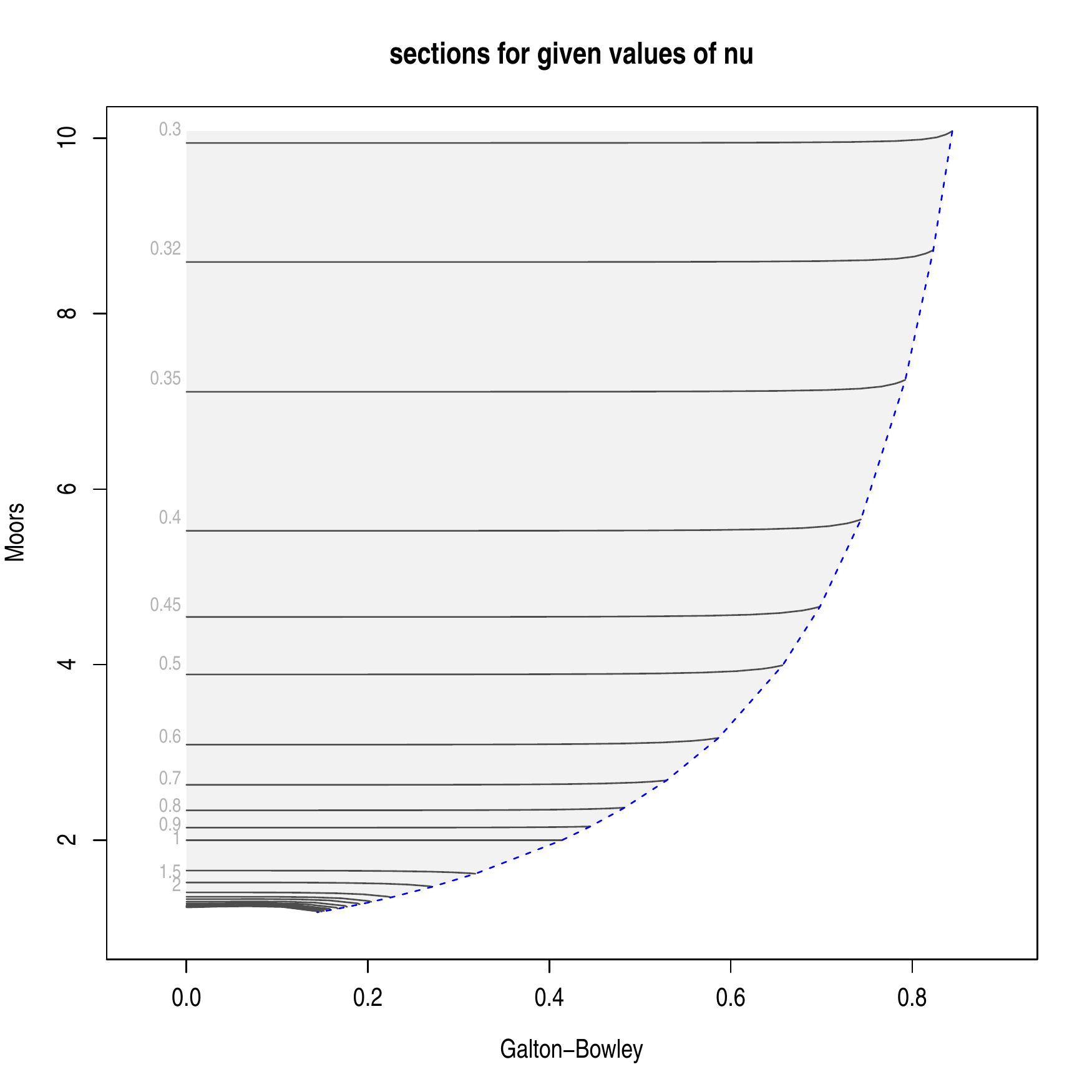}
\caption{\sf \textit{Loci} of the $(G,M)$ space for given values of $\delta$ 
  as $\nu$ varies (left plot) and for given values of $\nu$ as $\delta$ 
  varies (right plot).}
\label{f:(G,M)space-sections}
\end{figure}

Details of the numerical calculations are as follows.
The Galton-Bowley and the Moors measures have been evaluated over
a $13\times25$ grid of points identified by the selected values
\begin{eqnarray*}
 \delta^* &=& (0,~ 0.1,~  0.2,~  0.3,~  0.4,~  0.5,~  0.6,~  0.7,~  0.8,
         ~ 0.9,~  0.95,~ 0.99,~ 1), \\
 \nu^* &=& (0.30,~ 0.32,~ 0.35,~ 0.40,~ 0.45,~ 0.50,~  0.60,~ 0.70,~ 0.80,
          ~ 0.90,~ 1,~ 1.5,~  2,\\
     &&~ 3,~  4,~  5,~    7,~   10,~   15,~   20,~  30,~  40,~   50,
         ~  100,~  \infty)\,.
\end{eqnarray*}
The $G$ and $M$ values so obtained form the basis of 
Figure~\ref{f:(G,M)space-sections} and subsequent calculations.
Note that, while the vectors $\delta^*$ and $\nu^*$  identify 
a regular  grid of  points on the $(\delta, \nu)$ space,  
they corresponds to a curved grid  on the shaded regions of 
Figure~\ref{f:(G,M)space-sections}.

Boundary parameter values require a special handling.
Specifically, $\delta=1$ and correspondingly $\lambda=\infty$ identify
the Student's $t$ distribution truncated below 0, that is, the square
root transform of the Snedecor's $F(1,\nu)$ distribution; hence, in this
case the quantiles are computed as square roots of the $F(1,\nu)$ quantiles.
The associated points lie on the right-side boundary of the shaded area.
Another special value is $\nu=\infty$ which corresponds to the SN 
distribution.
In this case, function \texttt{qsn} of package \texttt{sn} has been 
used; the corresponding points lie on the concave curve in the bottom-left 
corner of the shaded area.

Conceptually, Figure~\ref{f:(G,M)space-sections} represents the key
for the inversion from $(G,M)$ to $(\delta, \nu)$ and equivalently
to $(\lambda, \nu)$ since $\lambda=(1-\delta^2)^{-1/2}\delta$.
However, in practical terms, we must devise a mechanism for inversely 
interpolating the $(G,M)$ values computed at the grid points.
Evaluation of this interpolation scheme at the sample values
$(\tilde{G}, \tilde{M})$ will yield the desired estimates. 

To this end, the second plot indicates the most favourable front 
for tackling the problem, since its almost horizontal lines show
that the Moors measure is very nearly a function of $\nu$ only.
Denote by $M^\circ$ the value of $M$  when $\delta=0$.
The 25 available values of $M^\circ$ at $\nu^*$ are reported in the second
column of Table~\ref{t:coef}; the remaining columns will be explained shortly.
From these 25 values, an interpolating  spline of  $1/\nu$  as a function 
of  $M^\circ$ has been  introduced. 
Use of the $1/\nu$ transformed variable  substantially reduces 
the otherwise extreme curvature of the function.
Evaluation of this spline function at the sample value $\tilde{M}$
and conversion into its reciprocal yields an initial estimate 
$\tilde\nu$.

\begin{table}[ht]
\centering
\caption{\sf Coefficients used for interpolation of the $G$ and $M$ tabulated values.}
\label{t:coef}
\small
\begin{tabular}{rrrrr}
  \hline
\rule[-3mm]{0mm}{7mm}\strut
% \multicolumn{1}{c}{$\nu$} 
$\nu^*$~~ &  $M\big|_{\delta=0} $ & $\eta^{(\nu)}_1$ & $\eta^{(\nu)}_2$ & $\eta^{(\nu)}_3$ \\ 
  \hline
  0.30 & 9.946 & 2.213831 & $-$0.315418 & $-$0.007641 \\ 
  0.32 & 8.588 & 2.022665 & $-$0.240821 & $-$0.012001 \\ 
  0.35 & 7.110 & 1.790767 & $-$0.164193 & $-$0.021492 \\ 
  0.40 & 5.525 & 1.506418 & $-$0.090251 & $-$0.047034 \\ 
  0.45 & 4.543 & 1.305070 & $-$0.050702 & $-$0.087117 \\ 
  0.50 & 3.888 & 1.156260 & $-$0.028013 & $-$0.143526 \\ 
  0.60 & 3.088 & 0.952435 & $-$0.005513 & $-$0.307509 \\ 
  0.70 & 2.630 & 0.819371 & 0.004209 & $-$0.536039 \\ 
  0.80 & 2.339 & 0.724816 & 0.008992 & $-$0.818739 \\ 
  0.90 & 2.142 & 0.653206 & 0.011596 & $-$1.142667 \\ 
  1.00 & 2.000 & 0.596276 & 0.013136 & $-$1.495125 \\ 
  1.50 & 1.652 & 0.417375 & 0.015798 & $-$3.365100 \\ 
  2.00 & 1.517 & 0.314104 & 0.016371 & $-$5.011929 \\ 
  3.00 & 1.403 & 0.192531 & 0.016274 & $-$7.304089 \\ 
  4.00 & 1.354 & 0.123531 & 0.015682 & $-$8.676470 \\ 
  5.00 & 1.327 & 0.080123 & 0.014987 & $-$9.546498 \\ 
  7.00 & 1.298 & 0.030605 & 0.013674 & $-$10.561206 \\ 
  10.00 & 1.277 & $-$0.003627 & 0.012113 & $-$11.335506 \\ 
  15.00 & 1.262 & $-$0.024611 & 0.010334 & $-$11.977601 \\ 
  20.00 & 1.254 & $-$0.030903 & 0.009149 & $-$12.343369 \\ 
  30.00 & 1.247 & $-$0.031385 & 0.007650 & $-$12.789281 \\ 
  40.00 & 1.244 & $-$0.027677 & 0.006721 & $-$13.074983 \\ 
  50.00 & 1.241 & $-$0.023285 & 0.006079 & $-$13.284029 \\ 
  100.00 & 1.237 & $-$0.005288 & 0.004478 & $-$13.874691 \\
   $\infty$ & 1.233 &  &  &  \\ 
  \hline
\end{tabular}
\end{table}
 
Consider now estimation of $\delta$ or equivalently of $\lambda$, a task
which essentially amounts to approximate the curves in the first plot
in Figure~\ref{f:(G,M)space-sections}.
After some numerical exploration, it turned out that a closely interpolating
function can be established in the following form:
\begin{equation}
  \log \lambda \approx \eta_1^{(\nu)}\,u + \eta_2^{(\nu)}\, u^3 + 
           \eta_3^{(\nu)}\, u^{-3}, \qquad\quad u=\log G.
  \label{eq:alpha(G)}         
\end{equation}
where the fitted values of the  coefficients  $\eta_j^{(\nu)}$ for the
selected $\nu^*$ values are reported in the last three columns of 
Table~\ref{t:coef}, with the exception of $\nu=\infty$.
Use of (\ref{eq:alpha(G)})  combined with the coefficients of Table~\ref{t:coef}
allows to find an approximate value of $\lambda$ for the selected values of
$\nu$. 
If an intermediate value of $\nu$ must be considered, such that
$\nu_1<\nu<\nu_2$ where $\nu_1, \nu_2$ are two adjacent values of $\nu^*$,
a linear interpolation of the corresponding coefficients is performed.
More explicitly, a value of $\eta_j^{(\nu)}$ is obtained by linear
interpolation of $\eta_j^{(\nu_1)}$ and $\eta_j^{(\nu_2)}$, for $j=1,2,3$;
then (\ref{eq:alpha(G)}) is applied using these interpolated coefficients.
If $\nu$ is outside the range of finite $\nu$ values in the first column
of Table~\ref{t:coef}, the $\eta_j^{(\nu)}$ values associated to the closest 
such value of $\nu$ are used. 

Operationally, we use the just-described scheme with $\nu$ set at the 
value $\tilde\nu$  obtained earlier, leading to an estimate 
$\tilde\lambda$ of $\lambda$.

Numerical testing of this procedure has been performed as follows. 
For a number of pairs of values $(\alpha,\nu)$, the corresponding octiles 
and  the $(G,M)$ measures have been computed and the proposed procedure 
has been applied to these measures.
The returned parameter values were satisfactorily close to the original 
$(\alpha,\nu)$  pair, with some inevitable discrepancies due to the 
approximations involved, but of limited entity. 
If necessary, a refinement can be obtained by numerical search targeted 
to minimize a suitable distance between a given  pair $(G,M)$  and 
the analogous values derived from the associated $(\alpha, \nu)$ pair. 
However, this refinement was not felt necessary in the numerical work 
described later, thanks to the good working of the above-described
interpolation scheme.

We are now left with estimation of $\xi$ and $\omega$. Bearing in mind the
representation $Y=\xi+\omega\,Z$ introduced just before (\ref{e:st-pdf-4par}),
$\omega$ is naturally estimated by
\begin{equation}
   \tilde\omega = \frac{q_3- q_1}{q^\ST_3- q^\ST_1}
   \label{e:tilde.omega}
\end{equation}
where the terms in the numerator are sample quartiles and those in the
denominator are quartiles of $Z\sim\ST(0,1, \tilde\lambda, \tilde\nu)$. 

Consideration  of $Y=\xi+\omega\,Z$ again says that an estimate of $\xi$ 
can be obtained as an adjustment of the sample median $q_2$ via
\begin{equation}
   \tilde\xi= q_2 - \tilde\omega\,q_2^\ST
   \label{e:tilde.xi}
\end{equation}
where $q_2^\ST$ is the median of $Z\sim\ST(0,1, \tilde\lambda, \tilde\nu)$. 

The estimates so produced are marked by an asterisk in the examples of 
Figure~\ref{f:deviance}, showing to perform well in those cases,
while requiring a negligible computing time compared to MLE.

%------------------------------------------------------------------------------
\subsection{Extension to the regression case}  \label{s:prelimFit-regres}

We want to extend the methodology of Section~\ref{s:prelimFit-core} 
to the regression setting where the location parameter varies across 
observations as a linear function  of a set of $p$, say, explanatory 
variables, which are assumed to include the constant term, 
as it is commonly the case.
If $x_i$ is the vector of covariates pertaining to the $i$th subject,
observation $y_i$ is now assumed to be drawn from 
$\ST(\xi_i, \omega, \lambda, \nu)$ where
\begin{equation}
  \xi_i = x_i\T\, \beta, \qquad i=1,\dots, n,
  \label{e:xi=Xbeta}
\end{equation}
for some $p$-dimensional vector $\beta$ of unknown parameters;
hence now the parameter vector is $\theta=(\beta\T, \omega, \lambda, \nu)\T$.
The assumption of independently drawn observations is retained.

The direct extension of the median as an estimate of location, 
which was used in  Section~\ref{s:prelimFit-core}, 
is an estimate of $\beta$  obtained by median regression,
which corresponds to adoption of the least absolute deviations 
fitting criterion instead of the more familiar least squares.
This can also be viewed as a special case of quantile regression,
when the quantile level is set at $1/2$.
A classical treatment of quantile regression is \cite{koenker:2005} and 
corresponding numerical work can be carried out using the \texttt{R} package 
\texttt{quantreg}, see \cite{koenker:2018quantreg}, among others tools.

Use of median regression delivers an estimate $\tilde\beta^m$ of $\beta$ 
and a vector of residual values, $r_i=y_i-x_i\T\tilde\beta^m$ 
for $i=1,\dots,n$.  
Ignoring $\beta$ estimation errors, these residuals are values sampled from 
$\ST(-m_0, \omega^2, \lambda, \nu)$, where $m_0$ is a suitable value, 
examined shortly, which makes the distribution to have 0 median,
since this is the target of the median regression criterion.
We can then use the same procedure of Section~\ref{s:prelimFit-core},
with the $y_i$'s replaced the $r_i$'s, to estimate $\omega, \lambda, \nu$,
given that the value of $m_0$ is irrelevant at this stage.

The final step is a correction to the vector $\tilde\beta^m$ to adjust
for the fact that $y_i-x_i\T\beta$ should have  median $m_0$, 
that is, the median of $\ST(0,\omega, \lambda, \nu)$, not median 0.
This amounts to increase all residuals by a constant value $m_0$,
and this step is accomplished by setting a vector $\tilde\beta$ with all
components equal to $\tilde\beta^m$ except that the intercept term, 
$\beta_0$ say,  is estimated by
\[
   \tilde\beta_0 = \tilde\beta^m_0 - \tilde\omega\, q_2^\ST 
\]
similarly to (\ref{e:tilde.xi}).

%------------------------------------------------------------------------------
\subsection{Extension to the multivariate case} \label{s:prelimFit-multiv}

Consider now the case of $n$ independent observations from a multivariate
$Y$ variable with density (\ref{e:mst-pdf}), hence 
   $Y\sim\ST_d(\xi, \Omega, \alpha, \nu)$.
This case can be combined with the regression setting of 
Section~\ref{s:prelimFit-regres}, so that the  $d$-dimensional location 
parameter varies for each observation  according to
\begin{equation}
  \xi_i\T = x_i\T\, \beta, \qquad i=1,\dots, n,
  \label{e:xi.d=Xbeta}
\end{equation}
where now $\beta=(\beta_{\cdot 1}, \dots, \beta_{\cdot d})$ is a 
$p\times d$ matrix of parameters.
Since we have assumed that the explanatory variables include a constant term, 
the regression case subsumes the one of identical distribution, when $p=1$.
Hence we deal with the regression case directly, where the $i$th observation
is sampled from $Y_i\sim \ST_d(\xi_i, \Omega, \alpha, \nu)$ and $\xi_i$ is
given by (\ref{e:xi.d=Xbeta}), for $i=1,\dots,n$.

Arrange the observed values in a $n\times d$ matrix $y=(y_{ij})$.  
Application of the procedure presented in Sections~\ref{s:prelimFit-core}
and~\ref{s:prelimFit-regres} separately to each column of $y$ delivers
estimates of $d$ univariate models. 
Specifically, from the $j$th column of $y$, we obtain estimates 
$\tilde\theta_j$ and corresponding `normalized' residuals $\tilde{z}_{ij}$:
\begin{equation}
 \tilde\theta_j=
   (\tilde\beta_{\cdot j}\T, \tilde\omega_j, \tilde\lambda_j, \tilde\nu_j)\T,
 \qquad
 \tilde{z}_{ij}= \tilde\omega_j\inv(y_{ij}- x_i\T \tilde\beta_{\cdot j}) \,.
 \label{e:Uv-fits}
\end{equation}
where it must be recalled that the `normalization' operation uses location
and scale parameters, but these do not coincide with the mean and the 
standard deviation of the underlying random variable.  

Since the meaning of expression (\ref{e:xi.d=Xbeta}) is to define 
a set of univariate regression modes with a common design matrix,
the vectors $\tilde\beta_{\cdot 1},\dots, \tilde\beta_{\cdot d}$ can 
simply be arranged in a $p\times d$ matrix $\tilde\beta$ which represents
an estimate of $\beta$.

The set of univariate estimates in (\ref{e:Uv-fits}) provide $d$ estimates
for $\nu$, while only one such a value enters the specification
of the multivariate ST distribution.
We have adopted the median of $\tilde\nu_1,\dots,\tilde\nu_d$ as the
single required estimate, denoted $\tilde\nu$.

The scale quantities $\tilde\omega_1, \dots, \tilde\omega_d$ estimate 
the square roots of the diagonal elements of $\Omega$, but off-diagonal 
elements require a separate estimation step. 
What is really required to estimate is the scale-free matrix $\bar\Omega$.
% since $\Omega_{ij}=\omega_i\bar\Omega_{ij}\omega_j$ and $\bar\Omega_{ii}=1$.
This is the problem examined next.

If $\omega$ is the diagonal  matrix formed by the squares roots of 
$\Omega_{11},\dots, \Omega_{dd}$, all variables  $\omega\inv(Y_i-\xi_i)$ 
have distribution  $\ST_d(0, \bar\Omega, \alpha, \nu)$, for $i=1,\dots,n$.
Denote by $Z=(Z_1, \dots,Z_d)\T$ the generic member of this set of variables. 
We are concerned with the distribution of the products $Z_jZ_k$, 
but for notational simplicity we focus on the specific product $W=Z_1\,Z_2$,
since all other products are of similar nature. 
 
We must then examine the distribution of $W{=}Z_1Z_2$ when $(Z_1, Z_2)$ 
is a bivariate ST variable. This looks at first to be a daunting
task, but a major simplification is provided by consideration of the
perturbation invariance property of symmetry-modulated distributions,
of which the ST is an instance. 
For a precise exposition of this property, see for instance Proposition~1.4 
of \cite{azza:capi:2014}, but in essence it states that, 
since $W$ is an even function of $(Z_1, Z_2)$, 
its distribution does not depend on $\alpha$, and it coincides with 
the distribution of the case $\alpha=0$, that is, the case of a
usual bivariate  Student's $t$ distribution, with dependence
parameter $\bar\Omega_{12}$.

Denote by $F_W(w;\rho,\nu)$ the distribution function of the product $W$ 
of variables $(Z_1, Z_2)$ having bivariate Student's $t$ density 
(\ref{e:mt-pdf}) in dimension $d=2$ with $\nu$ degrees of freedom 
and dependence parameter $\rho$ where $|\rho|<1$.
An expression of $F_W(w;\rho,\nu)$  is available in Theorem~1 of
\cite{wallgren:1980}. 
Although this expression involves numerical integration, this is not
problematic since univariate integration can be performed efficiently
and reliably with the \R\ function \texttt{integrate}. 

To estimate $\Omega_{12}$, we search for the value of $\rho$ such that
the median of the distribution of $W$ equates its sample value.
In practice, we compute the sample values  $\tilde{w}_1,\dots,\tilde{w}_n$ 
where $\tilde{w}_i=\tilde{z}_{i1}\tilde{z}_{i2}$, using the residuals in 
(\ref{e:Uv-fits}), and denote their median by $m_{\tilde{w}}$.
Then we must numerically solve the non-linear equation
\begin{equation}
   F_W(m_{\tilde{w}}; \rho, \tilde\nu) = 1/2
   \label{e:solveOmega12}
\end{equation}
with respect to $\rho$. 
Solution of the equation is facilitated by the monotonicity of 
$F_W(w; \rho, \nu)$ with respect to $\rho$, ensured by Theorem~2 of 
\cite{wallgren:1980}.
The solution of (\ref{e:solveOmega12}) is the estimate $\tilde\Omega_{12}$.

Proceeding similarly for other pairs of variables $(Z_j,Z_k)$, all entries
of matrix $\bar\Omega$ can be estimated; the diagonal elements are all~1. 
However, it could happen that the matrix so produced is not 
a positive-definite correlation matrix, because of estimation errors.
Furthermore an even more stringent condition has to be satisfied,  
namely that
\begin{equation}
  \Omega^* = \pmatrix{\bar\Omega & \delta \cr
                              \delta\T & 1 
                } > 0 \,.
  \label{e:Omega*>0}           
\end{equation}
This condition on $\Omega^*$ applies to all skew-elliptical distributions, 
of which the ST is an instance;
see \cite[p.\,101]{bran:dey:2001} or \cite[p.\,171]{azza:capi:2014}.

Denote by $\tilde\Omega^*$ the estimate of $\Omega^*$ with $\tilde{\bar\Omega}$ 
in  the $d\times d$ top-left block  obtained by solutions of equations of type 
(\ref{e:solveOmega12})  and $\tilde\delta$ computed by applying 
$\delta(\tilde\lambda_j)$ in (\ref{e:delta,b.nu}) to
$\tilde\lambda_1,\dots,\tilde\lambda_d$.
If  $\tilde\Omega^*$ is positive definite, then we move on to the next step; 
otherwise an adjustment is required.

There exist various techniques to adjust a nearly positive-definite matrix
to achieve positive-definiteness. Some numerical experimentation has been 
carried out using procedure \texttt{nearPD} of \R\ package \texttt{Matrix};
see \cite{bate:maec:2019Matrix}. 
Unfortunately, this did not work well when we used the resulting matrix 
for the next step, namely computation of the vector 
\begin{equation}
  \alpha = \big(1-\delta\T\bar\Omega\inv\delta\big)^{-1/2}\bar\Omega\inv\delta
  \label{e:delta->alpha}
\end{equation}
which enters the density function (\ref{e:mst-pdf}); 
see for instance equation (4) of \cite{azza:capi:2003}, which is stated for
the SN distribution, but it holds also for the ST.
The unsatisfactory outcome from  \texttt{nearPD} for our problem is 
presumably due to modifications in the relative size of the components 
of $\tilde\Omega^*$, leading to grossly inadequate $\alpha$ vectors, 
typically having a gigantic norm. 

A simpler type of adjustment has therefore been adopted, as follows.
If condition (\ref{e:Omega*>0}) does not hold for $\tilde\Omega^*$, 
the off-diagonal elements of the matrix are shrunk by a factor 0.95,
possibly repeatedly, until (\ref{e:Omega*>0}) is satisfied.
This procedure was quick to compute and it did not cause peculiar 
outcomes from (\ref{e:delta->alpha}). 

Hence, either directly from the initial estimates of $\bar\Omega$ and 
$\delta$ or after the adjustment step just described,
we obtain valid components satisfying condition (\ref{e:Omega*>0})
and a corresponding vector $\tilde\alpha$ from (\ref{e:delta->alpha}).
The final step is to introduce scale factors via
\[
   \tilde\Omega = \tilde\omega\,\tilde{\bar\Omega}\,\tilde\omega
\] 
where $\tilde\omega=\mathrm{diag}(\tilde\omega_1, \dots, \tilde\omega_d)$.
This completes estimation of $(\beta, \Omega, \alpha, \nu)$.

%------------------------------------------------------------------------------

\subsection{Simulation work to compare initialization procedures} 
\label{s:simul-init}

A number of simulation runs has been performed to examine the performance
of the proposed methodology.
The computing environment is \R\ version 3.6.0.
The reference point for these evaluations is the methodology
currently in use, as provided by the publicly available version of \R\ package 
\texttt{sn} at the time of writing, namely version 1.5-4; 
see \cite{azzalini:sn1.5-4}. 
This will be  denoted `the current method' in the following.
Since the role of the proposed method is to initialize the numerical MLE
search, not the initialization procedure \emph{per se}, we compare the
new and the current method with respect to final MLE outcome.
However, since the numerical optimization method used after 
initialization is the same, any variations in the results originate
from the different initialization procedures.

We stress again that in a vast number of cases the working of the current
method is satisfactory and we are aiming at improvements when dealing 
with `awkward samples'. 
These commonly arise with ST distributions having low degrees of freedom,
about $\nu=1$ or even less, but exceptions exist, such as the
second sample in Figure~\ref{f:deviance}.

The primary aspect of interest is improvement in the quality of data fitting.
This is typically expressed as an increase of the maximal achieved 
log-likelihood, in its penalized form.
Another desirable effect is improvement in computing time.

The basic set-up for such numerical experiments is represented by simple 
random samples, obtained as independent and identically distributed 
values drawn from a named $\ST(\xi, \omega, \lambda, \nu)$. 
In all cases we set $\xi=0$ and $\omega=1$.
For the other ingredients, we have selected the following values:
\begin{equation}
 \begin{array}{rcl}
  \lambda &:\quad & 0, \quad 2, \quad 8,\\
  \nu     &:\quad & 1,\quad  3,\quad  8, \\
    n     &:\quad&  50, ~ 100, ~ 250, ~ 500
 \end{array}
\label{e:simul-param}
\end{equation}
and, for each combination of of these values, $N=2000$ samples have been drawn.

The smallest examined sample size, $n=50$, must be regarded as a sort of
`sensible lower bound' for realistic fitting of flexible distributions
such as the ST. 
In this respect, recall the cautionary note of \cite[p.\,63]{azza:capi:2014} 
about the fitting of a SN distribution with small sample sizes.
Since the ST involves an additional parameter, notably one having strong effect
on tail behaviour, that annotation holds \emph{a fortiori} here.

For each of the  $3\times3\times4\times2000=72000$ samples so generated, estimation
of the parameters $(\xi, \omega, \lambda, \nu)$ has been carried out using 
the following methods.\par
\negindent M0: this is the current method, which maximizes 
   the penalized log-likelihood using  function \texttt{st.mple}  
   as described in Section~\ref{s:examples}.
\negindent M1: preliminary estimates are computed as described in 
   Section~\ref{s:prelimFit-core};
\negindent M2:  maximization  of the penalized log-likelihood, still using 
 function \texttt{st.mple}, but starting from the estimates of M1;
\negindent M3: similar to M2, but using a simplified form of M1, where
  only the location and scale parameters are estimated,
  setting $\lambda=0$ and $\nu=10$.
\par
\vspace{2ex}

An exhaustive analysis of the simulation outcome would be far too lengthy
and space consuming.
As already mentioned, our primary interest is on the differences of 
maximized log-likelihood. Specifically, if denote by $\log{\hat L_h}$
the maximized value of the penalized log-likelihood using method M$h$,
we focus on  the quantities $D_{20}$,  $D_{23}$ and  $D_{30}$,
where $D_{hk}=\log{\hat L_h}-\log{\hat L_k}$.
Table~\ref{t:diff.logL-simple} reports the observed frequencies of $D_{hk}$ 
values, grouped in  intervals
\[
   (-\infty,-20],~~ (-20,-2],~~  (-2,-0.2],~~ (-0.2,0],~~ (0,0.2],~~
           (0.2,2],~~ (2,20],~~ (20,\infty]
\]
crosstabulated either with $n$ or with $\nu$. 
%------------------------------------- 
\begin{table}
% summaries from file "Output/simul-UvS-2019-05-28_131638-summary.Rsave"
\caption{Frequency tables of grouped values of $D_{20}$, $D_{23}$ and $D_{30}$
  crossed with values of $n$ and of $\nu$ in case of simple random sampling.}
\label{t:diff.logL-simple}
\centering
Frequencies of $D_{20} \times n$\\[0.5ex]
\begin{tabular}{rrrrrrrrr}  
  \hline \multicolumn{1}{c}{$n\vphantom{^|_|}$}
 & \multicolumn{1}{c}{$(-\infty,-20]$} & \multicolumn{1}{c}{$(-20,-2]$} 
 & \multicolumn{1}{c}{$(-2,-0.2]$} & \multicolumn{1}{c}{$(-0.2,0]$} 
 & \multicolumn{1}{c}{$(0,0.2]$} & \multicolumn{1}{c}{$(0.2,2]$} 
 & \multicolumn{1}{c}{$(2,20]$} & \multicolumn{1}{c}{$(20, \infty]$} \\ 
  \hline
   50 &   1 &   0 &   5 & 9380 & 8562 &  47 &   3 &   2 \\ 
  100 &   0 &   0 &   0 & 9359 & 8606 &  31 &   3 &   1 \\ 
  250 &   0 &   0 &   0 & 9235 & 8730 &  24 &   5 &   6 \\ 
  500 &   0 &   0 &   0 & 9163 & 8807 &  10 &   8 &  12 \\ 
 total &  1 &   0 &   5 & 37137 & 34705 & 112 &  19 &  21 \\ 
   \hline
\end{tabular}
\par\vspace{3ex}
%----------------
 Frequencies of $D_{20} \times \nu$\\[0.5ex]
\begin{tabular}{rrrrrrrrr}
  \hline \multicolumn{1}{c}{$\nu\vphantom{^|_|}$}
 & $(-\infty,-20]$ & $(-20,-2]$ & $(-2,-0.2]$ & $(-0.2,0]$ & $(0,0.2]$ & 
                                   $(0.2,2]$ & $(2,20]$ & $(20, \infty]$ \\ 
  \hline
  1 &   1 &   0 &   1 & 12903 & 11035 &  21 &  18 &  21 \\ 
  3 &   0 &   0 &   2 & 11926 & 12054 &  17 &   1 &   0 \\ 
  8 &   0 &   0 &   2 & 12308 & 11616 &  74 &   0 &   0 \\ 
   \hline
\end{tabular}
%----------------
\par\vspace{3ex}
Frequencies of $D_{23} \times n$\\[0.5ex]
\begin{tabular}{rrrrrrrrr} 
  \hline \multicolumn{1}{c}{$n\vphantom{^|_|}$}
 & $(-\infty,-20]$ & $(-20,-2]$ & $(-2,-0.2]$ & $(-0.2,0]$ & $(0,0.2]$ & 
                                   $(0.2,2]$ & $(2,20]$ & $(20, \infty]$ \\ 
  \hline
   50 &   1 &   0 &   3 & 9445 & 8550 &   1 &   0 &   0 \\ 
  100 &   0 &   0 &   0 & 9324 & 8676 &   0 &   0 &   0 \\ 
  250 &   0 &   0 &   0 & 9117 & 8883 &   0 &   0 &   0 \\ 
  500 &   0 &   0 &   0 & 9011 & 8989 &   0 &   0 &   0 \\ 
 total &  1 &   0 &   3 & 36897 & 35098 &   1 &   0 &   0 \\ 
   \hline
\end{tabular}
%----------------
%\par\vspace{3ex}
%Frequencies of $D_{23} \times \nu$\\[0.5ex]
%\begin{tabular}{rrrrrrrrr}
%  \hline \multicolumn{1}{c}{$\nu\vphantom{^|_|}$}
% & $(-\infty,-20]$ & $(-20,-2]$ & $(-2,-0.2]$ & $(-0.2,0]$ & $(0,0.2]$ & 
%                                   $(0.2,2]$ & $(2,20]$ & $(20, \infty]$ \\ 
%  \hline
%  1 &   1 &   0 &   1 & 13374 & 10624 &   0 &   0 &   0 \\ 
%  3 &   0 &   0 &   1 & 11859 & 12140 &   0 &   0 &   0 \\ 
%  8 &   0 &   0 &   1 & 11664 & 12334 &   1 &   0 &   0 \\ 
%   \hline
%\end{tabular}
%----------------
\par\vspace{3ex}
Frequencies of $D_{30} \times n$\\[0.5ex]
\begin{tabular}{rrrrrrrrr} 
  \hline \multicolumn{1}{c}{$n\vphantom{^|_|}$}
 & $(-\infty,-20]$ & $(-20,-2]$ & $(-2,-0.2]$ & $(-0.2,0]$ & $(0,0.2]$ & 
                                   $(0.2,2]$ & $(2,20]$ & $(20, \infty]$ \\ 
  \hline
  0 &   0 &   0 &   3 & 8925 & 9020 &  47 &   3 &   2 \\ 
  100 &   0 &   0 &   0 & 9075 & 8890 &  31 &   3 &   1 \\ 
  250 &   0 &   0 &   0 & 9132 & 8833 &  24 &   5 &   6 \\ 
  500 &   0 &   0 &   0 & 9195 & 8775 &  10 &   8 &  12 \\ 
  tab0 &   0 &   0 &   3 & 36327 & 35518 & 112 &  19 &  21 \\ 
   \hline
\end{tabular}

\end{table}
%------------------------------------- 

We are not concerned with samples having values $D_{hk}$ in the interval 
$(-0.2, 0.2)$, since these differences are not relevant 
from an inferential viewpoint; 
just note that they constitute the majority of cases.
As for $D_{20}$, the fraction of cases falling outside $(-0.2, 0.2)$ is small,  
but it is not negligible, and this justifies our efforts to improve over M0. 
As expected, larger $D_{20}$ values occur more easily when $n$ or $\nu$ 
are small, but sometimes also otherwise.
In all but a handful of cases, these larger differences are on the 
positive sides, confirming the effectiveness of the proposed method
for initialization. 
The general indication is that both methods M2 and M3, 
and implicitly so M1, improve upon the current method M0.

Visual inspections of individual cases where M1 performs poorly 
indicates that the problem originates in the sample octiles, 
on which all the rest depends. 
Especially with very low $n$ and/or $\nu$, the sample octiles can
occasionally happen to behave quite differently from expectations, 
spoiling everything. 
Unfortunately, there is no way to get around this problem,
which however is sporadic.

Another indication from Table~\ref{t:diff.logL-simple} is that M3 is essentially
equivalent to M2, in terms of maximized log-likelihood, and in some cases
it is even superior, in spite of its simplicity.

Another aspect is considered is Table~\ref{t:diff.time} which reports 
computing times and their differences as frequencies of  time intervals. 
A value $t_k$ represents the computing time for estimation from 
a given sample using method M$k$, obtained by the first value reported  
by the \R\ function \texttt{system.time}. 
For M2 and M3, $t_k$ includes the time spent for initialization with M1,
although this is a very minor fraction of the overall time.
Clearly the samples considered are of quite different nature, especially
so for sample size. However, our purpose is solely comparative and, 
since exactly the same samples are processed  by the various methods, 
the comparison of average computing times is valid.
Table~\ref{t:diff.time} shows a clear advantage of M2 over M0 
in  terms of computing time and also some advantage, 
but less prominent, over M3.

%------------------------------------- 
 \begin{table}
\caption{Frequency table of computing time  and their  differences in the  
   case of simple random samples. The values of $t_2$ and $t_3$ include the
    time $t_1$ for  their initialization.}
\label{t:diff.time}
\centering
\begin{tabular}{crrrrrrrr}
  \hline $~\vphantom{^|_|}$
  & $(-\infty,-0.25]$ & $(-0.25,-0.1]$ & $(-0.1,-0.05]$ & $(-0.05,0]$ 
  & $(0,0.05]$ & $(0.05,0.1]$ & $(0.1,0.25]$ & $(0.25, \infty]$ \\ 
  \hline 
$t_0$ &   0 &   0 &   0 &   0 & 39561 & 21049 & 10815 & 575 \\ 
$t_1$ &   0 &   0 &   0 &   0 & 71998 &   0 &   2 &   0 \\ 
$t_2$ &   0 &   0 &   0 &   0 & 49086 & 20047 & 2866 &   1 \\ 
$t_3$ &   0 &   0 &   0 &   0 & 44178 & 21880 & 5942 &   0 \\ 
$t_2-t_0$ & 163 & 2010 & 4629 & 51101 & 13790 & 294 &  13 &   0 \\ 
$t_2-t_3$ &   0 &  30  & 1148 & 50694 & 19679 & 438 &  11 &   0 \\ 
$t_3-t_0$ & 121 & 1487 & 2783 & 46038 & 21451 & 116 &  4  &    0 \\
   \hline
\end{tabular}
\end{table}      
%------------------------------------- 

Additional simulations have been run having the location parameter
expressed via a linear regression.
Given  a vector $x$ formed by $n$ equally 
spaced points on the interval $(-1, 1)$, design matrices have been built
using $p$ transformations $T_j(x)$, inclusive of the constant function
$T_0(x)=1$, as  follows:
\[ 
 \begin{array}{rccccc} 
                         &\quad& T_0(x) ~ & T_1(x) & ~T_2(x) & ~T_3(x) \\
  \hbox{case A~} (p=3)   &:& 1 & x & ~\sqrt{1+x}  \\
  \hbox{case B~} (p=3)   &:& 1 & x & \sin 3x  \\
  \hbox{case C~} (p=4)   &:& 1 & x &  \sin 3x  & x/(1+0.8\,x)
 \end{array}
\]  
Computation of $T_j(x)$ over the $n$ values of $x$ yields the columns 
of the design matrix; the regression parameters $\beta_1,\dots,\beta_p$ 
have been set at $\beta_j=1$ for all $j$s.
For each of the A, B, C design matrices, and for each parameter 
combinations in (\ref{e:simul-param}), $N=2000$ have been generated,
similarly to the  case of simple random samples.

In Table~\ref{t:diff.logL-regress}, we summarize results only for case~C, 
as the other cases are quite similar. The distribution of $D_{20}$ in
the top two sub-tables still indicate a superiority of M2 over M0, 
although less pronounced than for simple samples.
The lower portion of the table indicates a slight superiority of M3 over
M2, reinforcing the similar indication from Table~\ref{t:diff.logL-simple}.
%------------------------------------- 
\begin{table}
% summaries from file "Output/simul-UvR-2019-05-05_103346-summary.Rsave"
\caption{Frequency tables of grouped values of $D_{20}$, $D_{23}$ and $D_{30}$
  crossed with values of $n$ and of $\nu$ in case of a linear regression setting
  with $p=4$ explanatory variables.}
\label{t:diff.logL-regress}
\centering
Frequencies of $D_{20} \times n$\\[0.5ex]
\begin{tabular}{rrrrrrrrr}  
  \hline \multicolumn{1}{c}{$n\vphantom{^|_|}$}
 & \multicolumn{1}{c}{$(-\infty,-20]$} & \multicolumn{1}{c}{$(-20,-2]$} 
 & \multicolumn{1}{c}{$(-2,-0.2]$} & \multicolumn{1}{c}{$(-0.2,0]$} 
 & \multicolumn{1}{c}{$(0,0.2]$} & \multicolumn{1}{c}{$(0.2,2]$} 
 & \multicolumn{1}{c}{$(2,20]$} & \multicolumn{1}{c}{$(20, \infty]$} \\ 
  \hline
   50 &   0 & 138 & 278 & 8480 & 8772 & 195 &  94 &  43 \\ 
  100 &   0 &  15 &  44 & 8829 & 8984 &  75 &  19 &  34 \\ 
  250 &   0 &   0 &   1 & 9029 & 8902 &  23 &   9 &  36 \\ 
  500 &   0 &   0 &   0 & 9326 & 8597 &  14 &  16 &  47 \\ 
  total &   0 & 153 & 323 & 35664 & 35255 & 307 & 138 & 160 \\  
   \hline
\end{tabular}
\par\vspace{3ex}
%----------------
 Frequencies of $D_{20} \times \nu$\\[0.5ex]
\begin{tabular}{rrrrrrrrr}
  \hline \multicolumn{1}{c}{$\nu\vphantom{^|_|}$}
 & $(-\infty,-20]$ & $(-20,-2]$ & $(-2,-0.2]$ & $(-0.2,0]$ & $(0,0.2]$ & 
                                   $(0.2,2]$ & $(2,20]$ & $(20, \infty]$ \\ 
  \hline
  1 &   0 & 118 & 188 & 12967 & 10291 & 156 & 120 & 160 \\ 
  3 &   0 &  23 &  88 & 11074 & 12759 &  44 &  12 &   0 \\ 
  8 &   0 &  12 &  47 & 11623 & 12205 & 107 &   6 &   0 \\ 
    \hline
\end{tabular}
%----------------
\par\vspace{3ex}
Frequencies of $D_{23} \times n$\\[0.5ex]
\begin{tabular}{rrrrrrrrr} 
  \hline \multicolumn{1}{c}{$n\vphantom{^|_|}$}
 & $(-\infty,-20]$ & $(-20,-2]$ & $(-2,-0.2]$ & $(-0.2,0]$ & $(0,0.2]$ & 
                                   $(0.2,2]$ & $(2,20]$ & $(20, \infty]$ \\ 
  \hline
  50 &   0 & 177 & 329 & 8413 & 8864 & 167 &  50 &   0 \\ 
  100 &   0 &  18 &  43 & 8802 & 9096 &  37 &   4 &   0 \\ 
  250 &   0 &   0 &   3 & 8899 & 9096 &   2 &   0 &   0 \\ 
  500 &   0 &   0 &   0 & 9066 & 8934 &   0 &   0 &   0 \\ 
  total &   0 & 195 & 375 & 35180 & 35990 & 206 &  54 &   0 \\ 
   \hline
\end{tabular}
%----------------
%\par\vspace{3ex}
%Frequencies of $D_{23} \times \nu$\\[0.5ex]
%\begin{tabular}{rrrrrrrrr}
%  \hline \multicolumn{1}{c}{$\nu\vphantom{^|_|}$}
% & $(-\infty,-20]$ & $(-20,-2]$ & $(-2,-0.2]$ & $(-0.2,0]$ & $(0,0.2]$ & 
%                                   $(0.2,2]$ & $(2,20]$ & $(20, \infty]$ \\ 
%  \hline
%  1 &   0 & 159 & 245 & 12582 & 10813 & 149 &  52 &   0 \\ 
%  3 &   0 &  24 &  89 & 11098 & 12758 &  29 &   2 &   0 \\ 
%  8 &   0 &  12 &  41 & 11500 & 12419 &  28 &   0 &   0 \\ 
%     \hline
%\end{tabular}
\par\vspace{3ex}
Frequencies of $D_{30} \times n$\\[0.5ex]
\begin{tabular}{rrrrrrrrr} 
  \hline \multicolumn{1}{c}{$n\vphantom{^|_|}$}
 & $(-\infty,-20]$ & $(-20,-2]$ & $(-2,-0.2]$ & $(-0.2,0]$ & $(0,0.2]$ & 
                                   $(0.2,2]$ & $(2,20]$ & $(20, \infty]$ \\ 
  \hline
  50 &   0 &  32 & 117 & 8856 & 8641 & 200 & 111 &  43 \\ 
  100 &   0 &   3 &  26 & 9069 & 8783 &  63 &  22 &  34 \\ 
  250 &   0 &   0 &   1 & 9160 & 8770 &  24 &   9 &  36 \\ 
  500 &   0 &   0 &   0 & 9188 & 8735 &  14 &  16 &  47 \\ 
  total &   0 &  35 & 144 & 36273 & 34929 & 301 & 158 & 160 \\ 
  \hline
   \end{tabular}
\end{table}
%------------------------------------- 

In the two subtables of Table~\ref{t:diff.logL-regress} about
$D_{20}$, note that there are 160 samples where M0 goes completely wrong. 
All these sample were generated with $\nu=1$, a fact which is not 
surprising considering the initial parameter selection of \texttt{st.mple}, 
in its standard working described at the beginning of Section~\ref{s:examples}. 
Since that initial selection is based on a least-squares fit of the regression 
parameters, this step clashes with the non-existence of moments when the 
underlying ST distribution has $\nu=1$ degrees of freedom. 
Not only the regression parameters are poorly fitted, but the ensuing 
residuals are spoiled, affecting also the initial fit of the other parameters.

A set of simulations  has also been run in the bivariate case, 
hence sampling from density (\ref{e:mst-pdf}) with $d=2$.
The scale matrix and the shape vector have been set to
\[
  \Omega= \pmatrix{ 1 & 1/2 \cr 1/2 & 1 }, \qquad
  \alpha = \lambda\: \pmatrix{ 1 \cr 2 }
\]
where $\lambda$ spans the values given in (\ref{e:simul-param}).
Also $n$ and $\nu$ have been set like in  (\ref{e:simul-param}), 
with the exception that $n=50$ has been not included, considering that 
$50$ data points would constitute a too small sample in the present context.
On the whole, $3^3\times 2000=54000$ bivariate samples have then
been generated. 
They have been processed by function \texttt{mst.mple} of package \texttt{sn} 
and the initialization method of Section~\ref{s:prelimFit-multiv}, 
with obvious modifications of the meaning of notation M0 to M3.

The summary output of the simulations is presented in Table~\ref{t:diff.logL-multiv}.
There is a clear winner this time, since M3 is constantly superior to the others.
Between M0 and M2, the latter is still preferable for $\nu=1$,
but not otherwise.

The almost constant superiority of M3 over M2 is quite surprising, 
given the qualitatively different indication emerging in the univariate case.
This rather surprising effect must be connected to transformation 
(\ref{e:delta->alpha}), as it has also been indicated by direct
examination of a number of individual cases: a moderate estimation 
error even of a single $\lambda_j$ component, and consequently
of $\delta_j$, transforms into a poor estimate of $\alpha$.
It so happens that the conservative choice $\alpha=0$ of M3
avoids problems and can be, in its simplicity, more effective.

%------------------------------------- 
\begin{table}
% summaries from file "Output/simul-MvS-2019-05-02_203537-summary.Rsave"
\caption{Frequency tables of grouped values of $D_{20}$, $D_{23}$ and $D_{30}$
  crossed with values of $n$ and of $\nu$ in the bivariate case.}
\label{t:diff.logL-multiv}
\centering
Frequencies of $D_{20} \times n$\\[0.5ex]
\begin{tabular}{rrrrrrrrr}  
  \hline \multicolumn{1}{c}{$n\vphantom{^|_|}$}
 & \multicolumn{1}{c}{$(-\infty,-20]$} & \multicolumn{1}{c}{$(-20,-2]$} 
 & \multicolumn{1}{c}{$(-2,-0.2]$} & \multicolumn{1}{c}{$(-0.2,0]$} 
 & \multicolumn{1}{c}{$(0,0.2]$} & \multicolumn{1}{c}{$(0.2,2]$} 
 & \multicolumn{1}{c}{$(2,20]$} & \multicolumn{1}{c}{$(20, \infty]$} \\ 
  \hline
  100 &  28 &  71 & 238 & 9410 & 8169 &  61 &   9 &  14 \\ 
  250 &   8 &   8 &  29 & 9613 & 8326 &   7 &   1 &   8 \\ 
  500 &   0 &   3 &   4 & 9330 & 8657 &   1 &   1 &   4 \\ 
 total &  36 &  82 & 271 & 28353 & 25152 &  69 &  11 &  26 \\ 
   \hline
\end{tabular}
\par\vspace{3ex}
%----------------
 Frequencies of $D_{20} \times \nu$\\[0.5ex]
\begin{tabular}{rrrrrrrrr}
  \hline \multicolumn{1}{c}{$\nu\vphantom{^|_|}$}
 & $(-\infty,-20]$ & $(-20,-2]$ & $(-2,-0.2]$ & $(-0.2,0]$ & $(0,0.2]$ & 
                                   $(0.2,2]$ & $(2,20]$ & $(20, \infty]$ \\ 
  \hline
  1 &   1 &   3 &   8 & 8368 & 9578 &  14 &   2 &  26 \\ 
  3 &  12 &  14 &  61 & 9814 & 8094 &   5 &   0 &   0 \\ 
  8 &  23 &  65 & 202 & 10171 & 7480 &  50 &   9 &   0 \\ 
  \hline
\end{tabular}
%----------------
\par\vspace{3ex}
Frequencies of $D_{23} \times n$\\[0.5ex]
\begin{tabular}{rrrrrrrrr} 
  \hline \multicolumn{1}{c}{$n\vphantom{^|_|}$}
 & $(-\infty,-20]$ & $(-20,-2]$ & $(-2,-0.2]$ & $(-0.2,0]$ & $(0,0.2]$ & 
                                   $(0.2,2]$ & $(2,20]$ & $(20, \infty]$ \\ 
  \hline
  100 &  28 &  71 & 232 & 9607 & 8006 &  50 &   6 &   0 \\ 
  250 &   8 &   8 &  29 & 9747 & 8205 &   3 &   0 &   0 \\ 
  500 &   0 &   3 &   4 & 9611 & 8382 &   0 &   0 &   0 \\ 
  total &  36 &  82 & 265 & 28965 & 24593 &  53 &   6 &   0 \\ 
  \hline
\end{tabular}
%----------------
%\par\vspace{3ex}
%Frequencies of $D_{23} \times \nu$\\[0.5ex]
%\begin{tabular}{rrrrrrrrr}
%  \hline \multicolumn{1}{c}{$\nu\vphantom{^|_|}$}
% & $(-\infty,-20]$ & $(-20,-2]$ & $(-2,-0.2]$ & $(-0.2,0]$ & $(0,0.2]$ & 
%                                   $(0.2,2]$ & $(2,20]$ & $(20, \infty]$ \\ 
%  \hline
%  0 &  16 &  10 &  85 & 11986 & 5898 &   5 &   0 &   0 \\ 
%  4 &  14 &  49 &  60 & 7997 & 9871 &   8 &   1 &   0 \\ 
%  16 &   6 &  23 & 120 & 8982 & 8824 &  40 &   5 &   0 \\ 
%     \hline
%\end{tabular}
%
\par\vspace{3ex}
Frequencies of $D_{30} \times n$\\[0.5ex]
\begin{tabular}{rrrrrrrrr}
  \hline \multicolumn{1}{c}{$n\vphantom{^|_|}$}
 & $(-\infty,-20]$ & $(-20,-2]$ & $(-2,-0.2]$ & $(-0.2,0]$ & $(0,0.2]$ & 
                                   $(0.2,2]$ & $(2,20]$ & $(20, \infty]$ \\ 
  \hline
  100 &   0 &   0 &  25 & 8924 & 8998 &  36 &   3 &  14 \\ 
  250 &   0 &   0 &   2 & 9035 & 8946 &   8 &   1 &   8 \\ 
  500 &   0 &   0 &   0 & 8839 & 9155 &   1 &   1 &   4 \\ 
 total &   0 &   0 &  27 & 26798 & 27099 &  45 &   5 &  26 \\ 
     \hline
\end{tabular}

\end{table}
%------------------------------------------------------------------------------
\subsection{Conclusions}

The overall indication of the simulation work is that the proposed preliminary
estimates work quite effectively, providing an improved initialization  
of the numerical MPLE search. 
The primary aspect is that higher log-likelihood values are usually achieved,
compared to the currently standard method, M0, sometimes by a remarkable 
margin. Another positive aspect is the saving in the overall computing time.

Of the two variant forms of the new initialization, leading to methods 
M2 and M3, the latter has emerged as clearly superior in the multivariate case,
but no such clear-cut conclusion can be drawn in the univariate setting,
with indications somewhat more favourable for M2.
In this case, it is advisable to consider both variant of the preliminary 
estimates and carry out two numerical searches. 
Having to choose between them, the quick route is to take the one
with higher log-likelihood. 
However, direct inspection of both outcomes must be recommended,
including exploration of the profile log-likelihood surface.

Surely, it would have been ideal to identify a universally superior method, 
to be adopted for all situations, but this type of simplification still 
eludes us.

%-----------------------------------------------------------------------------    -

\FloatBarrier
%\input{references}
%\bibliography{strings,sn+related,misc-distn,books,thispaper} 

\end{document}